\documentclass[a4paper,fleqn,usenatbib]{mnras}

\usepackage[T1]{fontenc}
\usepackage{ae,aecompl}

\usepackage{graphicx}	
\usepackage{amsmath}	
\usepackage{amssymb}	
\usepackage{color}
\usepackage{aas_macros}
\usepackage{longtable}

\title[Circumstellar disc sizes in Be/X-ray binaries]{On the relationship between circumstellar disc size and X-ray outbursts in \bf{Be/X-ray binaries}}
\author[I. Monageng et al.]
  {I. M. Monageng$^{1,2}$\thanks{E-mail: itu@saao.ac.za}, 
  V.A. McBride$^{1,2}$
  M.J. Coe$^3$,
  I.A. Steele$^4$
  and P. Reig$^{5,6}$
  \\
$^1$South African Astronomical Observatory, P.O Box 9, Observatory, 7935, Cape Town, South Africa\\
$^2$Department of Astronomy, University of Cape Town, Private Bag X3, Rondebosch 7701, South Africa\\
$^3$School of Physics and Astronomy, University of Southampton, Highfield, Southampton SO17 1BJ\\
$^4$Astrophysics Research Institute, Liverpool John Moores University, L3 5RF, UK\\
$^5$IESL, Foundation for Research and Technology-Hellas, GR-71110 Heraklion, Crete, Greece\\
$^6$Department of Physics \& Institute of Theoretical Computational Physics, University of Crete,\\ PO Box 2208, GR-710 03, Heraklion, Crete, Greece
}

\date{Accepted 2016 September 13. Received 2016 September 13; in original form 2016 July 25}

\pubyear{2015}

\def\LaTeX{L\kern-.36em\raise.3ex\hbox{a}\kern-.15em
    T\kern-.1667em\lower.7ex\hbox{E}\kern-.125emX}

\def\g09{G10}

\def\lsim{\mathrel{\hbox{\rlap{\hbox{\lower4pt\hbox{$\sim$}}}\hbox{$<$}}}}
\def\gsim{\mathrel{\hbox{\rlap{\hbox{\lower4pt\hbox{$\sim$}}}\hbox{$>$}}}}

\makeatletter
\newcommand*{\rom}[1]{\expandafter\@slowromancap\romannumeral #1@}
\newcommand*{\Rom}[1]{\expandafter\@slowromancap\romannumeral #1@}
\makeatother

\begin{document}

\label{firstpage}
\pagerange{\pageref{firstpage}--\pageref{lastpage}}
\maketitle

\begin{abstract}
We present long term H$\alpha$ monitoring results of five Be/X-ray binaries to study the Be disc size variations and their influence on Type \rom{2} (giant) X-ray outbursts. The work is done in the context of the viscous decretion disc model which predicts that Be discs in binary systems are truncated by resonant torques induced by the neutron star in its orbit. Our observations show that type \rom{2} outbursts are not correlated(nor anti-correlated) with the disc size, as they are seen to occur both at relatively small and large Be disc radii. We discuss these observations in context of alternate interpretation of Be disc behaviour, such as precession, elongation and density effects, and with cognisance of the limitations of our disc size estimates. \end{abstract}


\section{Introduction}
\label{sec:introduction}

Be/X-ray binaries (BeXBs) are binary stellar systems which make up the largest subclass of the high mass X-ray binary (HMXB) population (about two-thirds of the identified systems, \citealt{2011BASI...39..429P}). These systems are primarily composed of a massive, early-type donor (an Oe or Be star) and a neutron star accretor. The rapidly rotating Oe/Be primary is surrounded by a geometrically thin, Keplerian circumstellar disc in the equatorial regions. The presence and variations of the disc are detected observationally through infrared excess and Balmer emission lines in the optical spectra, the strongest and best-studied of these being the H$\alpha$ emission line \citep{2003PASP..115.1153P}. The neutron star is generally in an eccentric orbit ($e \geq 0.3$) around the massive companion and a correlation in the $P_\mathrm{orb} - P_\mathrm{spin}$ diagram is observed, across a wide range of orbital periods (24.3 $\leq P_\mathrm{orb} \leq$ 262.6 days, \citealt{Corbet1984,Corbet1986}). The interaction of the neutron star with the material in the circumstellar disc results in transient X-ray behaviour. The transient nature of BeXBs is characterised by two types of outburst events \citep{1986ApJ...308..669S}:\\
\indent (\Rom{1}) Type \rom{1} (normal outbursts), which have moderate luminosities ($L \leq 10^{37} \mathrm{erg\,s^{-1}}$) and occur regularly, separated by the orbital period.\\
\indent (\rom{2}) Type \Rom{2} (giant outbursts), which display larger luminosities ($L \geq 10^{37} \mathrm{erg\,s^{-1}}$). They are less frequent and are not modulated on the orbital period.

The extent to which the circumstellar disc grows is restricted by the neutron star in its orbit. The viscous decretion disc model explains how circumstellar discs interact with the neutron star, resulting in disc truncation \citep{1991MNRAS.250..432L, 2001A&A...377..161O}. Angular momentum is added to the inner regions of the disc by the Be star and the material is then transported to the outer parts via viscosity (Shakura-Sunyaev $\alpha$-prescription for viscous stress is applied: \citealt{ShakuraSunyaev1973}). The disc is truncated at particular radii, where the ratio between the angular frequency of the disc rotation and that of the orbit is an integer. In BeXBs, the maximum observed H$\alpha$ equivalent width correlates with the orbital period \citep{ReigFabregatCoe1997,ReigNersesianZezas2016}, and this is interpreted as evidence for truncation of the circumstellar disc at a radius resonant with the neutron star orbit. This truncation is an expected outcome of the viscous disc model \citep{1991MNRAS.250..432L, OkazakiBateOgilvie2002} and its effect on mass transfer has been explored by \citet{NegueruelaOkazaki2001} and \citet{NegueruelaOkazakiFabregat2001}. Except for this tidal truncation, the neutron star, being much less massive than the Be star, has very little effect on the circumstellar disc.  However, when the circumstellar disc is truncated at a radius close to the inner Lagrangian point at periastron mass transfer to the neutron star can occur, resulting in an X-ray outburst (see Fig. 6 in \citealt{NegueruelaOkazaki2001} for an illustration of the geometry). 

This truncation suggests that the inner parts of Be star discs may be vertically thicker and/or denser than those of their isolated Be star counterparts, while the outer disc density drops much more rapidly \citep{OkazakiBateOgilvie2002}. While the physical extent of a Be star disc is very difficult to determine, the H$\alpha$ emission line in Be stars is often used to provide an estimate. The Balmer lines are optically thick and produced through recombination. They are also formed in a large part of the disc, giving a much better idea of the physical extent than some of the helium or metal lines which are formed closer to the central star. Balmer line equivalent widths have been shown to correlate with estimates from optical interferometry \citep{2006ApJ...651L..53G}. \citet{1972ApJ...171..549H} show that the peak separation in double-peaked emission lines resulting from a disc can be used to gauge the size of an emitting region, providing reliable sizes in the case of optically thin discs \citep{Hummel1994}. Interferometric measurements suggest that disc emitting areas in the various bands are as large as a few tens of stellar radii, with disc sizes larger in the $K$ band -- in rough agreement with the dynamic estimates from Huang \citep{2007ApJ...654..527G}.

\citet{SilajJonesTycner2010} show the dependence of Be star line profiles and equivalent widths on inclination angles, density and different power law indices of the radial density dependence. In these models the H$\alpha$ equivalent width correlates inversely with the peak separation, meaning both of these parameters are sensitive to changes in the disc size and/or density.

While the work of \citet{2006ApJ...651L..53G} and \citet{SilajJonesTycner2010} pertains to isolated Be stars, it is not clear how this translates to a case where the disc is truncated. \citet{ZamanovReigMarti2001} show a rough inverse correlation between H$\alpha$ equivalent width and peak separation for Be stars in BeXBs, and this seems to suggest that the situation is not so different in the case of the truncated disc. However, a degeneracy between the disc density and disc radius appears to persist. 

Our goal in this paper is to investigate how the circumstellar disc size variations, through studying different properties of the H$\alpha$ emission line, influence X-ray outbursts in BeXBs in the context of the viscous decretion disc model. The objects under study are Galactic Be/X-ray transients \hbox{1A~0535+262}, \hbox{4U~0115+634}, \hbox{V~0332+53}, \hbox{EXO~2030+375} and \hbox{1A~1118$-$61}.

The article is structured as follows: in section \ref{sec:observations} we present the observations and analysis of the data, while our method of estimating the circumstellar disc radius is addressed in section~\ref{sec:radius}. The results are presented in section~\ref{sec:results} and implications of the results are discussed in section~\ref{sec:discussion}.

\section{Observations}
\label{sec:observations}

\subsection{Optical observations}

As part of a long term monitoring programme of Galactic BeXBs we have collected optical spectra between 1 January 2009 and 21 July 2013 with the Liverpool Telescope \citep{2001AN....322..307S} and the Southern African Large Telescope (SALT, \citealt{2006IAUS..232....1B}). These are supplemented with data from a number of other telescopes: 1.0-m Jacobus Kapteyn Telescope (La Palma), 2.5-m Isaac Newton Telescope (La Palma), 1.3-m Skinakas Telescope (Crete), 4.2-m William Herschel Telescope (La Palma), South African Astronomical Observatory 1.9-m telescope (South Africa). 

For the SALT observations the Robert Stobie Spectrograph (RSS) was used in longslit mode. The PG2300 grating (lines ruled at 2300 l\,mm$^{-1}$) was used with a wavelength coverage $\lambda\lambda6200-7000\enskip$\AA\ at exposure times of 180\,s. A slit width of 1.5$\arcsec$ was used for the observations, with a resolution of $\sim$1.6 \AA\ and a resulting dispersion of 0.27 \AA /pix.  Wavelength calibration of SALT data was performed by fitting lines in the observed arc with those of reference SALT spectra using a Neon lamp. The data reduction process of the semi-reduced SALT data (SALT pipeline performs overscan correction, bias subtraction, gain correction and corrections for amplifier cross-talk, \citealt{2012ascl.soft07010C}) was undertaken with version 2.15.1 of IRAF\footnote{Image Reduction and Analysis Facility: iraf.noao.edu}. Flux calibration was not applied, since the aim of obtaining the spectra was to measure $EW(H\alpha)$ of emission lines as well as to locate the peaks of double-peaked H$\alpha$ emission lines.

On the Liverpool Telescope (LT) the Fibre-fed Robotic Dual-beam Optical Spectrograph (FRODOSpec) integral field unit spectrograph was used in conjunction with the red Volume Phase Holographic grating. This covered a wavelength range $\lambda\lambda5900-8000\enskip$\AA\  at exposure times of $200$\,s (1A~0535+262) and $600$\,s (4U~0115+63 and V~0332+53), resulting in a central resolution of 1.3\,\AA\ and a dispersion of 0.6\,\AA$/$pix.  The reduction of raw data obtained with FRODOSpec was performed through a fully autonomous pipeline \citep{2012AN....333..101B}. This comprised fibre tramline map generation, extraction of flux, arc fitting, sky subtraction and throughput correction.

1A~0535+262, 4U~0115+634, and V~0332+53 were also observed from the 1.3-m 
telescope of the Skinakas observatory located in the island of Crete 
(Greece). The spectra were obtained between the period 1999--2013. The 
telescope was equipped with a 2000 $\times$ 800 15 $\mu$m square pixel ISA SITe CCD and a 
1302\,l\,mm$^{-1}$ grating, giving a nominal dispersion of 1.04 \AA/pix
and a field of view of 19.6 arcmin $\times$ 7.8 arcmin.

Spectra were also obtained from the Anglo Australian, Jacobus Kapteyn, Isaac Newton and William Herschel telescopes, and the South African Astronomical Observatory. Details of these data and their reductions have been previously published in \citet{HaighCoeFabregat2004} and \citet{CoeBirdHill2007}. 

The equivalent width and peak separation measurements utilised in this work are presented in Appendix~\ref{append}.

\subsection{X-ray data}
X-ray lightcurves from the \emph{RXTE} All Sky Monitor (ASM) and the \emph{Swift} BAT Hard X-ray monitor have been used alongside the optical data as a tracer of the outburst and quiescent behaviour in the sources under study. 

\textit{RXTE} was active between December 1995 and January 2012. The All Sky Monitor (ASM, \citealt{LevineBradtCui1996}) aboard \emph{RXTE} scanned approximately 80\% of the sky every 90 minutes. The ASM lightcurves used in this work were downloaded from the archive\footnote{http://xte.mit.edu/ASM\_lc.html} and comprised one-day averages of the source count rate in the  2--10\,keV energy band.

Since 2004, the launch date of the Swift observatory, we also use X-ray lightcurves from the Burst Alert Telescope (BAT, \citealt{BarthemlyBarbierCummings2005}). The BAT observes 88\% of the sky every day, and the daily average lightcurves in the 15 -- 50\,keV band are available through the webpages of the Swift/BAT Hard X-ray transient monitor\footnote{http://swift.gsfc.nasa.gov/results/transients/} \citep{KrimmHollandCorbet2013}.

We perform a multiplicative conversion of the BAT count rates to equivalent ASM count rates by using the overlapping data (i.e. from 2004 to 2012). The scaling has been done so that outbursts and base flux levels match in these instruments.  These lightcurves are plotted in units of luminosity in 2--10\,keV band in the lower panels of the figures in section \ref{sec:results}. The distances used for the luminosities are presented in Table~\ref{tab:bex}. This is not a rigorous scaling (which would account for the X-ray spectrum across the energy range) but it allows us to distinguish Type I and Type II outbursts.

There are no easily accessible X-ray lightcurves for epochs prior to 1995.  X-ray all sky monitoring was possible between 1991 and 1998 through BATSE on CGRO, and through the Ariel All Sky Monitor between 1974 and 1980. So for epochs prior to December 1995, no X-ray lightcurves are shown, but Type II outbursts as reported in the literature are indicated as vertical lines on the figures.

\section{Circumstellar disc radius}
\label{sec:radius}

\begin{table*}
\label{tab:bex}   
   \caption{BeXBs and their parameters.  $v\sin i$ is the projected rotational velocity of the donor star, $M_*$ is the mass of the donor star, $i$ is the estimated inclination of the orbit to the plane of the sky, $P_\mathrm{orb}$ is the orbital period and $e$ the orbital eccentricity. The mean $V$-band magnitude is provided in the last column. }
   \begin{tabular}{ | l c c c c c c c | p{6cm} |}
   \hline
    Object & $v \sin i$ (km\,s$^{-1}$)$^\ddag$ & $M_{\ast}$ (M$_{\odot}$)$^{*}$& $i^{\dag}$ & $P_\mathrm{orb}$(d) &  $e^\star$ & $d$(kpc)$^{\dag\dag}$ &$V$ mag$^{\oplus}$ \\ \hline
    \hline
    1A~0535+262 & 254 & 20 & 27$^\circ$ & 111.1 & 0.47 & $\sim$2 & 9.39 \\ 
    4U~0115+63 & 365 & 18 & 43$^\circ$  & 24.32 & 0.34 & 7.8 & 15.19 \\
    V~0332+53 & 145 & 20 & 10.3$^\circ$   & 34.67 & 0.37 & $\sim$7 & 15.13\\
    EXO~2030+375 & 295 & 23 & 56$^\circ$  & 46.02 & 0.416 & $\sim$5 & 19.7 \\ 
    1A~1118$-$61 & 300 & 18 & 25$^\circ$  & 24.00 & 0.0 & 5 & 12.12 \\ \hline
    \end{tabular}\\
    \small{$\ddag$ \mbox{\citet{1998MNRAS.297..657C}}, \mbox{\citet{HutchingsCrampton1981}}, \mbox{\citet{1988A&A...202...81J}}, \mbox{\citet{NegueruelaOkazaki2001}}, \mbox{\citet{JanotPacheco1981}} ; $^*$\mbox{\citet{NegueruelaOkazaki2001}}, \mbox{\citet{NegueruelaOkazaki2001}}, \mbox{\citet{MotchPakullJanot1988}}; $^\dag$\mbox{\citet{CoeReigMcBride2006}}, \mbox{\citet{NegueruelaOkazaki2001}}, \mbox{\citet{CoePayneHanson1987}}, \mbox{\citet{ReigStevensCoe1998}}, \mbox{\citet{2001A&A...377..161O}}; $^\star$ \mbox{\citet{1994AIPC..308..459F}};$\dag\dag$\mbox{\citet{1998MNRAS.297L...5S}}, \mbox{\citet{NegueruelaOkazaki2001}}, \mbox{\citet{1999MNRAS.307..695N}}, \mbox{\citet{1989ApJ...338..359P}}, \mbox{\citet{JanotPacheco1981}} ;$^\oplus$\mbox{\citet{2003AJ....125.2531R}}}   
    
\end{table*}

\citet{1972ApJ...171..549H} demonstrated that the size of the H$\alpha$ emitting region of the circumstellar disc can be estimated by using the peak separation, $\Delta V$, of the double peaked H$\alpha$ emission lines if Keplerian velocity distribution of matter in the disc is assumed. We start from Newton's form of Kepler's third law, and taking peak separation of the double peaked H$\alpha$ emission line as given by $\Delta V = 2v_\mathrm{pv},$ where $v_\mathrm{pv}$ is the projected velocity at the outer edge of the H$\alpha$ emitting part of the disc. Taking into consideration the geometry of the disc with respect to our line of sight, the true velocity of the disc is then given by $v_\mathrm{tr} = v_\mathrm{pv}/\sin i$, where $i$ is the inclination angle of the disc. The true velocity of the disc can therefore be expressed by $v_{tr} = \Delta V/2\sin i$. Assuming that the orbits of particles in the disc are circular, with the prevailing force being the gravity of the Be star, the radius of the H$\alpha$ emitting region of the disc can be given by:
\begin{equation} 
	r = \frac{GM_{*} \sin^{2}i}{(0.5\Delta V)^2}  \centering
 \end{equation}
For single peaked profiles the peak separation, $\Delta V$, cannot be measured directly. However, \citet{1989Ap&SS.161...61H} demonstrated that there exists a linear relationship between $\Delta V$ and $EW(H\alpha)$:
 \begin{equation} 
	 \log\left(\frac{\Delta V}{2v \sin i}\right) = -a \log \left(\frac{-EW}{\textrm{\AA}}\right) + b
 \end{equation}

\begin{figure}
	\includegraphics[width = 0.45\textwidth]{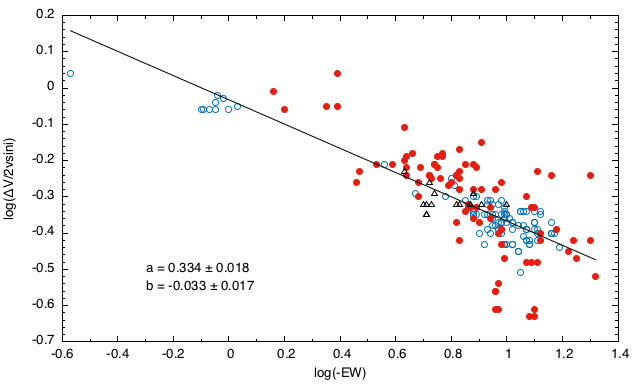}
	\caption{A plot of peak separation and EW(H$\alpha$) measurements of 1A~0535+262 (\textcolor{cyan}{$\circ$}), 4U~0115+634 (\textcolor{red}{$\bullet$}) and V~0332+53 ($\triangle$). The best fit line is over plotted, with fit parameters, $a$ and $b$.}
	\label{fit}
\end{figure}

From a fit to the $ \log(\frac{\Delta V}{2v \sin i})$ vs. $\log (\frac{-EW}{\textrm{\AA}})$ plot for double-peaked line profiles, the values of the constants $a$ and $b$ can be obtained from the slope and intercept, respectively. So for emission lines which display single peaked profiles the disc radius could be estimated using measurements of the EW(H$\alpha$). \citet{ZamanovReigMarti2001} derived parameters $a$ and $b$ for a sample of stars in a comparison study between circumstellar discs in BeXBs and isolated Be stars. This method of disc size estimation was applied to 1A~0535+262 (\citealt{CoeReigMcBride2006}) and more recently to HMXBs LSI+61$^\circ$303, MWC 148 and MWC 656 (\citealt{2016arXiv160505811Z}). \\
In our sample of BeXBs, only 1A~0535+262, 4U~0115+634 and V~0332+53 showed double peaked H$\alpha$ during some observations. Table~1 lists the parameters $M_{\ast}$, $v\sin i$ and $i$ obtained from literature, and the fit for these sources is shown in Fig.~\ref{fit}. 

The errors in the radius calculations were propagated from the errors in $\Delta V$ and EW(H$\alpha$). The typical error of a radius estimate obtained from direct measurements of $\Delta V$ (i.e double-peaked H$\alpha$ lines) has a range 10-15$\%$. For single-peaked profiles the radius error bars display a larger range of 50-70 $\%$ the radius value -- mainly due to the large scatter in the plot of $ \log(\frac{\Delta V}{2v \sin i})$ vs. $\log (\frac{-EW}{\dot{\mathrm{A}}})$, Fig.~\ref{fit}. The indirect estimates of $\Delta V$ (from inference using the relationship between $\Delta V$ and EW(H$\alpha$)) were checked with the double-peaked H$\alpha$ profiles and the two methods give radius estimates that agree within error bars.

\section{Results}
\label{sec:results}

\subsection{1A~0535+262}
1A~0535+262 is one of the best studied X-ray binary transients, discovered by \textit{Ariel} \rom{5} during a type \rom{2} outburst \citep{1975Natur.256..628R}. The system consists of an O9.7\rom{3}e primary star \citep{1980A&AS...40..289G,1998MNRAS.297L...5S} and an X-ray pulsar in orbit around the primary. 1A~0535+262 is at a distance $\sim2$ kpc \citep{1998MNRAS.297L...5S}. Orbital measurements of the binary system reveal an orbital period of $\sim111.1$ days \citep{1994AIPC..308..459F, 2006HEAD....9.0759F}, an eccentricity of $\sim 0.47$ (Finger et al. 1994) and pulse period of 103.39\,s \citep{2007AA...465L..21C}. 

\begin{figure*}
	\includegraphics[width = 0.9\textwidth]{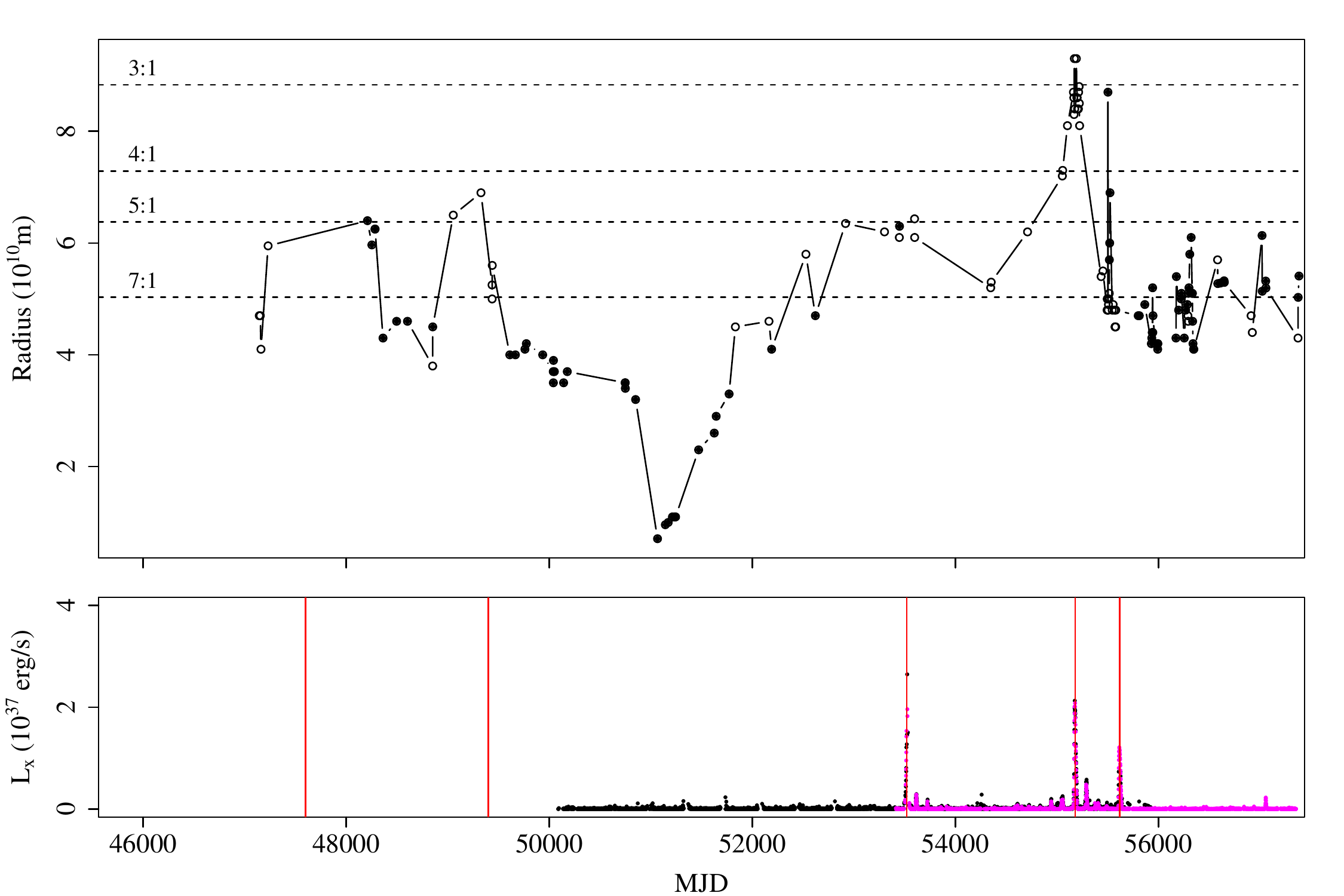}
	\caption{Top panel: Evolution of the Be disc in 1A~0535+262. The size of the 3:1 $r_\mathrm{res}$, 4:1 $r_\mathrm{res}$, 5:1 $r_\mathrm{res}$ and 7:1 $r_\mathrm{res}$ are indicated in the plot. The filled circles indicate the radius calculations from double-peaked profiles of H$\alpha$ and open circles represent single-peaked profiles. Lower panel: Long term X-ray lightcurve from the RXTE \textit{ASM} and Swift \textit{BAT}. The vertical lines in this panel show the peak times of Type II X-ray outbursts.}
	\label{0535}
\end{figure*}

During a period when there were no recorded X-ray outbursts ($\sim$MJD\,47140 -- $\sim$MJD\,47232, Fig.~\ref{0535}) the disc radius was below the 5:1 $r_\mathrm{res}$, with radii ranging from $r \sim 4.1 \times 10^{10}$\,m to $r \sim 6.2 \times 10^{10}$\,m. 1A~0535+262 underwent a type \rom{2} outburst which peaked at $\sim$MJD\,47600 (March 1989, \citealt{GiovannelliGraziati1992}) for which there was no optical coverage of the system. Between MJD\,48209 and MJD\,49054 the disc displayed sizes below the 4:1 $r_\mathrm{res}$ ranging from $r \sim 3.8 \times 10^{10}$\,m to $r \sim 6.7 \times 10^{10}$\,m unaccompanied by X-ray activity (BATSE all sky monitoring started around MJD~48350). A type \rom{2} outburst was then observed, peaking at at $\sim$MJD\,49400 and lasted for $\sim$50\,days with a peak flux of $\sim$8\,Crab at 20-50\,keV \citep{1997ApJS..113..367B}. During this period the disc size was still below the 4:1 $r_\mathrm{res}$, with a maximum radius of $r \sim 6.9 \times 10^{10}$\,m (MJD\,49327).

The disc appeared to have reached stability, with truncation at the 7:1 $r_\mathrm{res}$ for almost 3.5 years ($\sim$MJD\,49611 -- $\sim$MJD\,50850 ) before going through a disc-loss phase as evidence by H$\alpha$ in absorption (MJD\,51055). A similar trend in the disc size was found by \cite{2007ApJ...660.1398G} in their study of 1A~0535+262 at this epoch. 1A~0535+262 underwent its first outburst after a 10 year period of quiescence which peaked at $\sim$MJD\,53500 (May/June 2005) and reached a flux of 4.5\,Crab at 30\,keV \citep{2005ATel..557....1S}, followed by two type \rom{1} outbursts. A series of type \rom{1} outbursts starting at $\sim$MJD\,54610. The fourth of these was unusual in that it was double-peaked, with the first peak occurring $\sim$14 days before periastron passage and the second peak around periastron passage, at $\sim$MJD\,55058.

At the start of the LT monitoring programme (MJD~55163) the disc radius as inferred from H$\alpha$ measurements was close to 3:1 $r_\mathrm{res}$ ($r \sim 8.7 \times 10^{10}$\,m). During this period 1A~0535+262 was already undergoing a type \rom{2} outburst which peaked at $\sim$MJD\,55180 with a peak flux of $\sim$5.14\,Crab at 15--50\,keV \citep{2010arXiv10032969C}. Following was a series of type \rom{1} outbursts, the second of which was peculiar in that it was double-peaked. The disc size had declined to truncation below the 5:1 $r_\mathrm{res}$ at $\sim$MJD\,55436 while 1A~0535+262 underwent another type \rom{2} outburst in February 2011 (peaking at $\sim$MJD\,55620). 1A~0535+262 has since been in quiescence following the type \rom{2} outburst in 2011 with the disc truncated below the 5:1 $r_\mathrm{res}$.

\subsection{4U~0115+63}
4U~0115+63 is another well studied BeXB transient. The system was initially reported as a discovery by the $Uhuru$ satellite during its 1971 type \rom{2} outburst \citep{1972ApJ...178..281G, 1978ApJS...38..357F}. However, it was later noticed, after going through the $VELA$ 5B archival data that it had already been detected in 1969 \citep{1989ApJ...338..381W}. The optical companion in the binary system is a B0.2\rom{5}e star at a distance of 7.8 kpc (Negueruela \& Okazaki 2001). Using the \textit{Burst and Transient Source Experiment (BATSE)} monitoring data of a sample of accreting pulsars, \citet{1997ApJS..113..367B} derived the orbital parameters: $P_\mathrm{orb} = 24.317037(2)$\,d, $a_{X}\sin i = 140.13(8)$\,lt-s, $e = 0.3402(2)$, $T_\mathrm{peri}$ = MJD\,49279.2677(34). The pulse period of the pulsar is 3.614690(2)\,s \citep{1992ApJ...389..676T}. The Type I outbursts in 4U~0115+63 are well explained in context of the truncated viscous decretion disc model and the particular orbital geometry of this source, while the semi-regular Type II outbursts are speculated to occur due to a warped, precessing circumstellar disc \citep{NegueruelaOkazakiFabregat2001}.

For the earliest measurements ($\sim$MJD\,47936) the disc radius was comparable to periastron passage of the NS. During this period 4U~0115+63 underwent a type \rom{2} outbursts in February 1990 ($\sim$MJD\,47900, \citealt{1992ApJ...389..676T}). In March 1991 ($\sim$MJD\,48320) the system underwent another type \rom{2} outburst (\citealt{CominskyRobertsFinger1994}), with the radius just above periastron passage a month before the peak of the outburst. A type \rom{2} outburst was again observed in May/June 1994 ($\sim$MJD\,49500, \citealt{NegueruelaGroveCoe1997}) with no optical coverage. Following this the disc appeared to have undergone a disc-growth phase, reaching a maximum of $r \sim 7.7 \times 10^{10}$\,m at $\sim$MJD\,50094 and then declining to a radius below near the 5:1 $r_\mathrm{res}$ ($\sim$MJD\,50480). A type \rom{2} outburst, coinciding with this growth phase, was observed in November/December 1995 (peaking at $\sim$MJD\,50040) with a peak flux of 0.7 \,Crab at 3--300\,keV \citep{1995IAUC.6272....2S}, as well as a rare series of type \rom{1} outbursts separated by the orbital period (between MJD\,50300 and MJD\,50400) during the disc decline. 
Two type \rom{2} outbursts were again observed, peaking at $\sim$MJD\,51250 (March 1999) and $\sim$MJD\,51750 (August 2000) when the apparent disc size reached values above periastron passage, up to $r \sim 1.5 \times 10^{11}$\,m ($\sim$MJD\,51743). Following the August 2000 outburst, the disc underwent a low state, with radius values significantly below the 5:1 $r_\mathrm{res}$ (between $r \sim 7 \times 10^{9}$\,m and $r \sim 2.2 \times 10^{10}$\,m). The disc recovered to reach radii above periastron passage ($r \sim 1.4 \times 10^{10}$\,m, $\sim$MJD\,53180) on a time-scale of $\sim$260 days, accompanied by a type \rom{2} outburst peaking at $\sim$MJD\,53260. The disc decreased in size following the outburst but still above periastron passage, reaching radii up to $r \sim 7 \times 10^{10}$\,m with no X-ray activity. 4U~0115+63 underwent a further type \rom{2} outburst which peaked at $\sim$MJD\,54560 for which there is no simultaneous optical coverage.

\begin{figure*}
	\centering
	\includegraphics[width=0.85\textwidth]{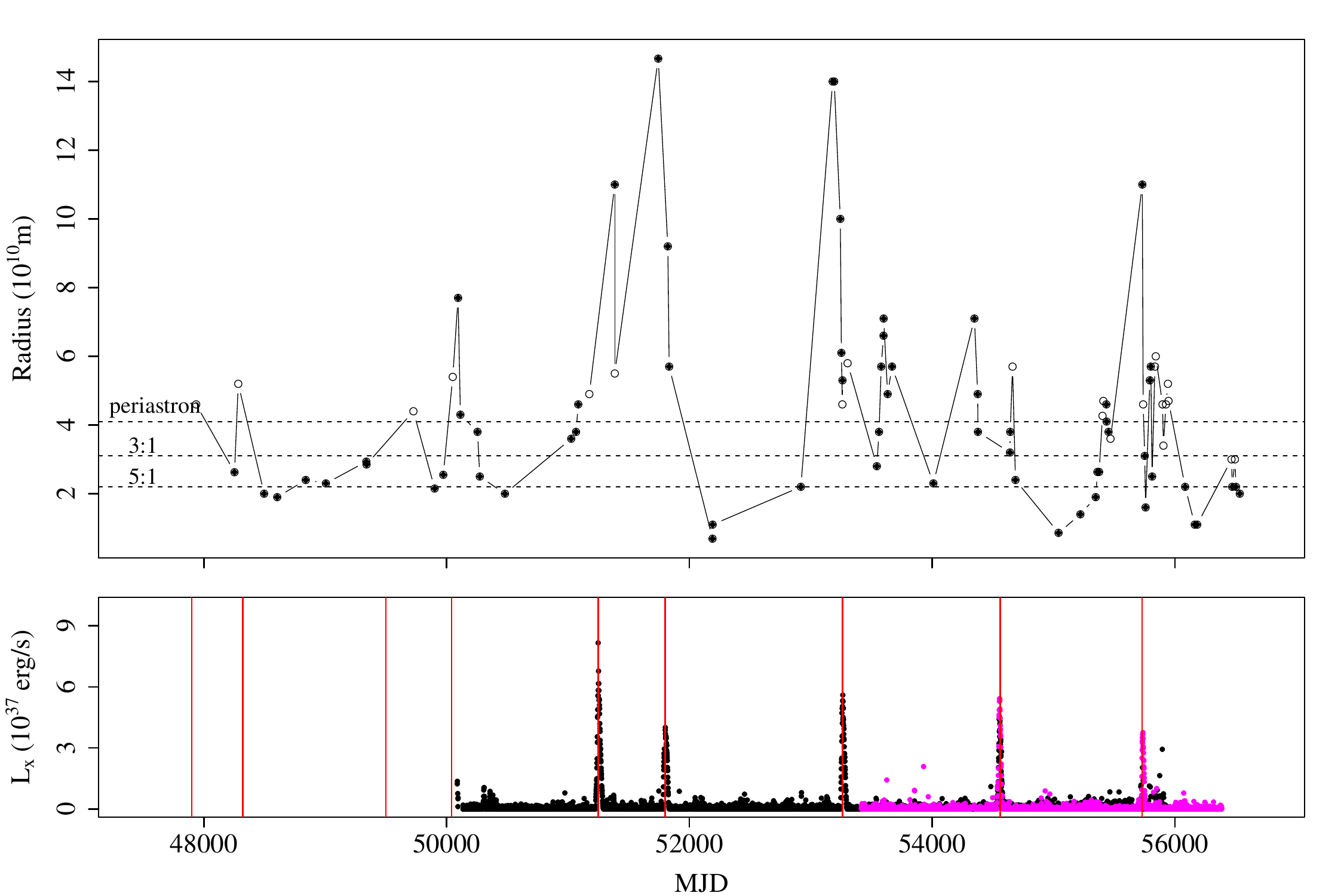}
	\caption{Top panel: Evolution of the Be disc in 4U~0115+63. The sizes of the 3:1 $r_\mathrm{res}$, 5:1 $r_\mathrm{res}$ and periastron distance of the neutron star are indicated in the plot. The filled circles indicate the radius calculations from double-peaked profiles of H$\alpha$ and open circles represent single-peaked profiles. Lower panel: Long term X-ray lightcurve from the RXTE \textit{ASM} and Swift \textit{BAT}. The vertical lines in this panel show the peak times of Type II X-ray outbursts.}
	\label{0115X}
\end{figure*}

At the beginning of the LT monitoring of 4U~0115+63 (MJD~55220) the disc displayed sizes below the 5:1 $r_\mathrm{res}$, at $r \sim 1.4 \times 10^{10}$\,m and grew to sizes larger than periastron passage, peaking at $\sim$MJD\,55731 ($r \sim 1.1 \times 10^{11}$\,m). A type \rom{2} outburst was observed during disc-growth, peaking at $\sim$MJD\,55730 (May/June 2011), which reached a flux of $\sim$0.8\,Crab at energy band 3--300\,keV \citep{BoldinTsygankovLutovinov2013}. The disc has since returned to a low state, with sizes below the 3:1 $r_\mathrm{res}$ and has been in quiescence since the type \rom{2} outburst in May/June 2011.

\subsection{V~0332+53}
V~0332+53 is a transient X-ray pulsar which was discovered by $VELA$ 5B during an outburst in 1973 \citep{1984ApJ...285L..15T}. The optical companion of the system is an O8-9\rom{5}e star, BQ Cam \citep{1999MNRAS.307..695N}, located at a distance $\sim$7 kpc. Using data from Rossi X-ray Timing Explorer (\textit{RXTE}) and \textit{INTEGRAL} observations, \citet{2005ApJ...630L..65Z} derived the following parameters: $P_\mathrm{spin} \approx 4.375$\,s, $P_\mathrm{orb} = 34.67$\,d, $e = 0.37$, $a_{X} \sin i \approx 86$\, lt-s.

\begin{figure*}
	\centering
	\includegraphics[width = 0.85\textwidth]{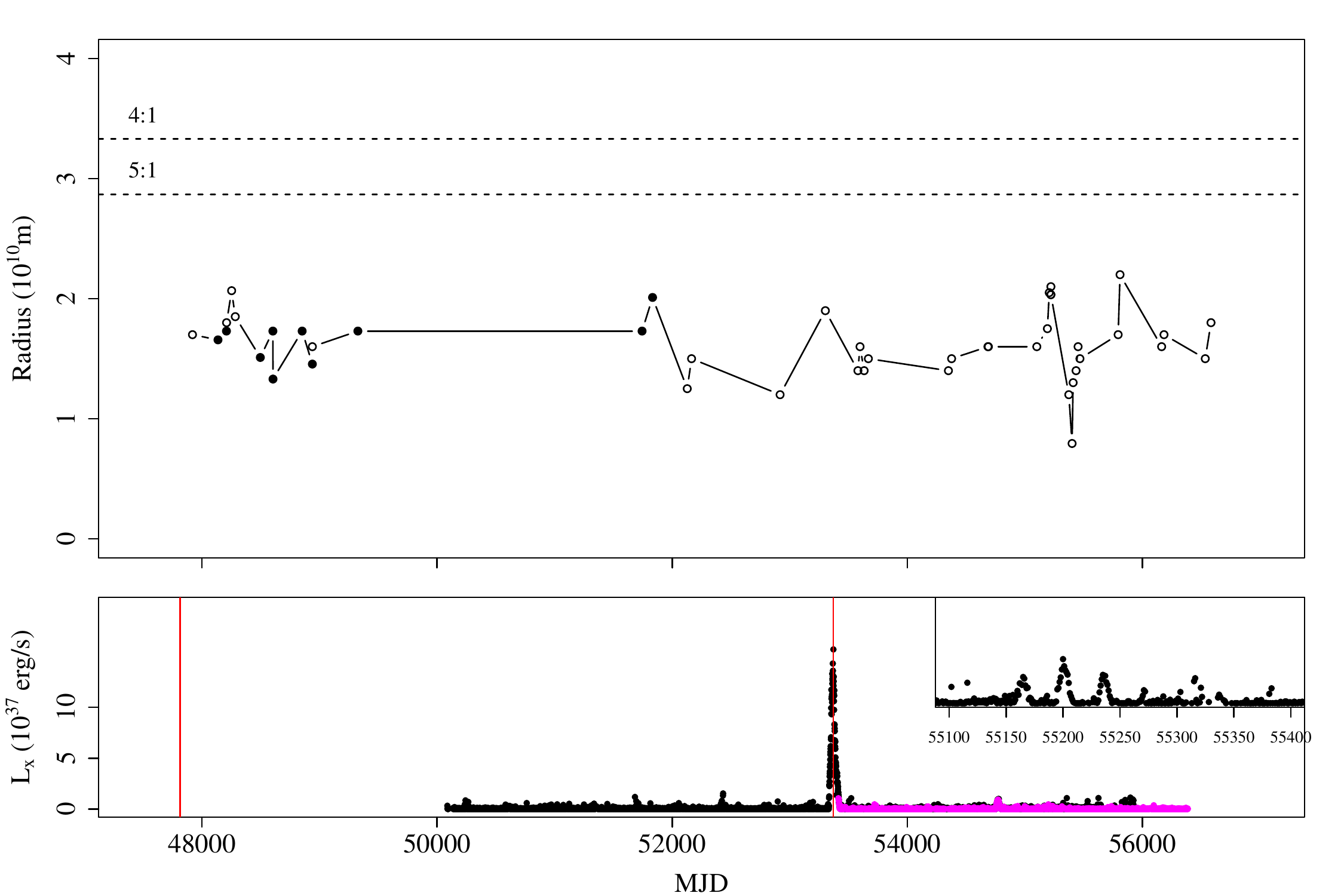}
	\caption{Top Panel: Evolution of the Be disc in V~0332+53. The size of the 4:1 $r_\mathrm{res}$ and 5:1 $r_\mathrm{res}$ are indicated in the plots. The filled circles indicate the radius calculations from double-peaked profiles of H$\alpha$ and open circles represent single-peaked profiles. Lower panel: Long term X-ray lightcurve from the RXTE \textit{ASM} and Swift \textit{BAT}. The vertical lines in this panel show the peak times of Type II X-ray outbursts. The inset is a zoomed portion of the X-ray lightcurve illustrating the series of Type I outbursts in 2009.}
	\label{v0332X}
\end{figure*}

The apparent disc size in V~0332+53 has remained below the 5:1 $r_\mathrm{res}$ since the beginning of the optical monitoring. Prior to the commencement of the optical observations, V~0332+53 had already undergone a type \rom{2} outburst which reached a peak flux of $\sim$0.3\,Crab (2--20\,keV) peaking at $\sim$MJD 47814 \citep{MakishimaMiharaIshida1990}. The system underwent two X-ray outburst events for which there is no simultaneous optical coverage: a type \rom{2} outburst occurred between November 2004 and February 2005 (peaking at $\sim$MJD\,53370) and a type \rom{1} outburst which peaked at $\sim$MJD\,54770.
%

The LT monitoring of V~0332+53 started when the system was undergoing a rare series of type \rom{1} outbursts in 2009 ($\sim$MJD\,55160 -- $\sim$MJD\,55320, see inset on Fig.~\ref{v0332X}). The disc was truncated below the 5:1 $r_\mathrm{res}$ during this period. V~0332+53 has been in quiescence since the end of the 2009 series of type \rom{1} outbursts. 

\subsection{EXO~2030+375}

EXO~2030+375 is an X-ray transient pulsar which was discovered with the \textit{European X-ray Observatory Satellite} (EXOSAT) during a giant outburst in 1985 (MJD 46203) \citep{1989ApJ...338..359P}. The system is located at a distance of 5.3\,kpc \citep{1989ApJ...338..359P}. Optical and infrared studies of the system revealed the optical companion to be an early-type main-sequence star (B0\rom{5}e), with properties similar to those of V~0332+53 \citep{1987A&A...182L..55M, 1988MNRAS.232..865C}. In the analysis of long-term \textit{RXTE} data, \citet{2005ApJ...620L..99W} obtained orbital parameters: $P_\mathrm{orb} = 46.0202(2)$\,d, $e = 0.416(1)$, $a_X\sin i = 238(2)$\,lt-s, $T_\mathrm{peri}$ = JD\,2451099.93(2). The spin period of the pulsar is 41.4106(1)\,s \citep{2013ApJ...764..158N}. 

 \begin{figure*}
	\includegraphics[width = 0.85\textwidth]{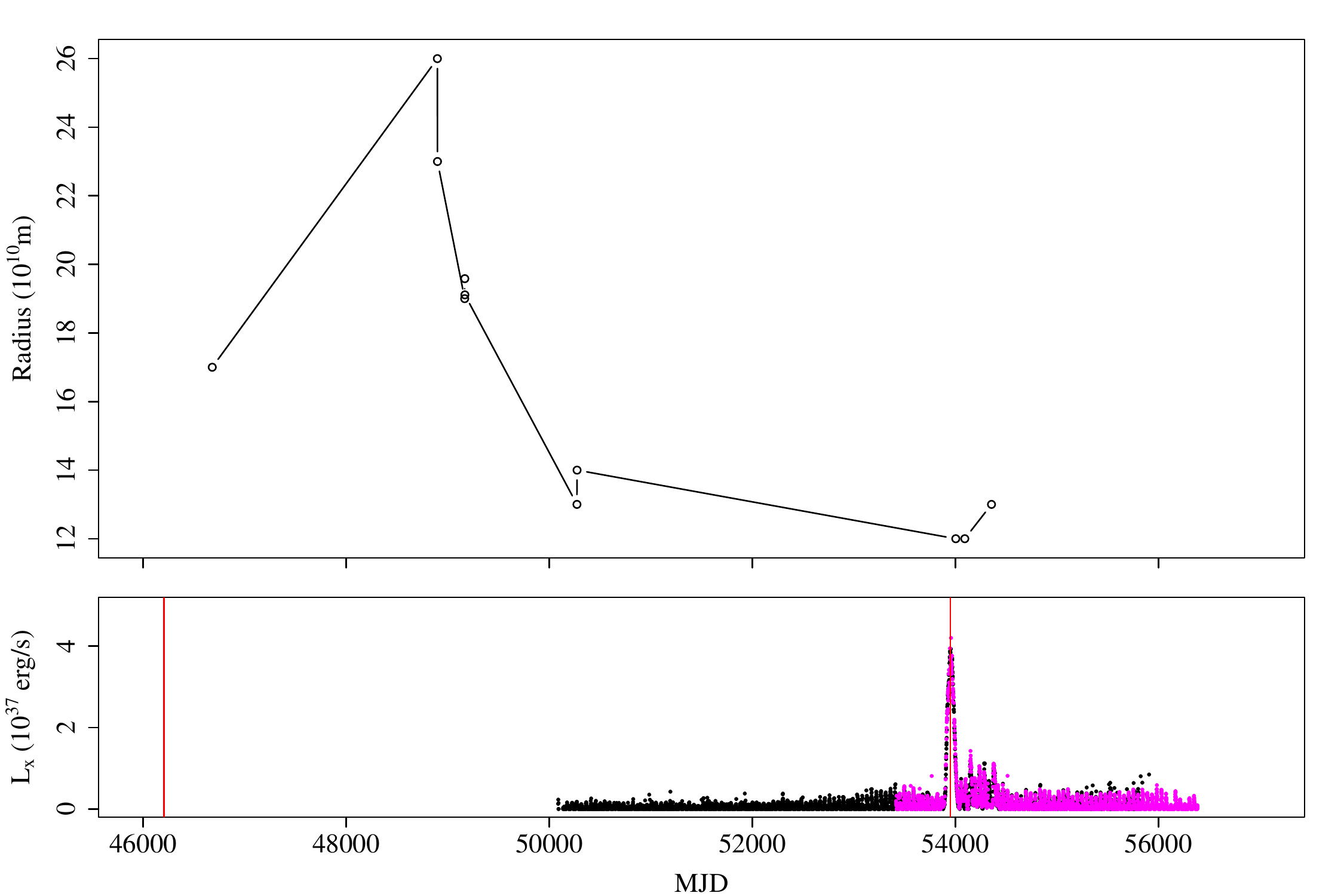}
	\caption{Top panel: Evolution of the Be disc in EXO~2030+375. Lower panel: Long term X-ray lightcurve from the RXTE \textit{ASM} and Swift \textit{BAT}. The vertical lines in this panel show the peak times of Type II X-ray outbursts. The vast number of Type I outbursts is clearly visible throughout the X-ray lightcurve. The 5:1, 4:1, 3:1 resonance radii and periastron passage, not included in the plot on this scale, range between $r \sim 3.5 \times 10^{10}$m and $r \sim 6.1 \times 10^{10}$m.}
	\label{ON2030}
\end{figure*}

Between MJD\,48897 and MJD\,49168 the radius of the Be disc in EXO~2030+375 showed a range of values all greater than the neutron star distance periastron, from $r \sim 1.5 \times 10^{11}$\,m to $r_\mathrm{res}$, at $r \sim 2.1 \times 10^{11}$\,m (Fig.~\ref{ON2030}). The disc size had declined to $r \sim 1.2 \times 10^{11}$\,m by MJD\,50273. During this period EXO~2030+375 was displaying low intensity type \rom{1} outbursts \citep{ReigStevensCoe1998}. The system underwent a rare type \rom{2} outburst which peaked at $\sim$MJD\,53950, with the disc size at $r \sim 1.2 \times 10^{11}$\,m on the outburst decline but showing an increasing trend. \\
 EXO~2030+375 is unusual among BeXBs in that it goes through long periods of type \rom{1} outbursts that last for a few years. Other BeXBs, in contrast, undergo series of type \rom{1} outbursts that usually last for a few orbits of the NS. The fact that the disc size is often larger than periastron distance could explain this (the NS interacts with the disc). This would support the idea that type I outbursts occur because the NS physically interacts with the disc at or near periastron. When the radius of the disc is smaller or of the order of periastron passage, X-ray activity decreases because the NS no longer passes through the disc.

\subsection{1A~1118$-$61}
1A~1118$-$61 is an X-ray transient which was discovered by the \textit{Ariel-5} satellite in 1974 during a type \rom{2} outburst \citep{1975IAUC.2752....1E}. The optical companion in the BeXB is an O9.5\rom{4}-\rom{5}e star at a distance $5\pm2$\,kpc \citep{JanotPacheco1981}. In the analysis of data obtained with \textit{RXTE/PCA}, \citet{2011A&A...527A...7S} derived the orbital parameters: $P_\mathrm{orb} = 24.0\pm0.4$\,d, $a_{X}\sin i = 54.85\pm1.4$\,lt-s, $e = 0.0$, $P_\mathrm{spin} = 407.6546\pm0.0011$\,s. 

 \begin{figure*}
	\centering
	\includegraphics[width = 0.85\textwidth]{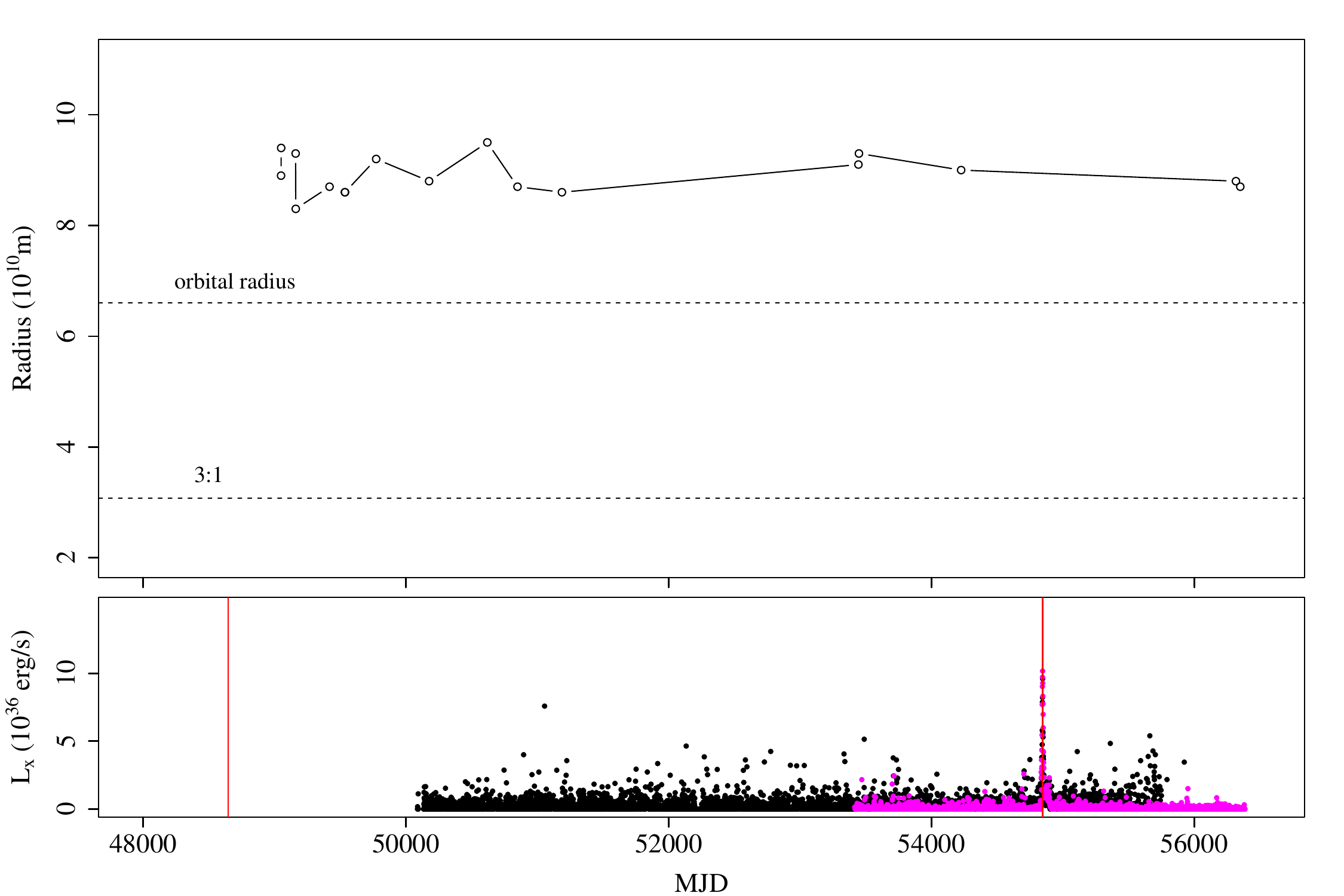}
	\caption{Top Panel: Evolution of the Be disc in 1A~1118$-$61. The 3:1 resonance radius and the orbital radius of the neutron star are indicated. Lower panel: Long term X-ray lightcurve from the RXTE \textit{ASM} and Swift \textit{BAT}. The vertical lines in this panel show the peak times of Type II X-ray outbursts.} 
	\label{1118rad}
\end{figure*}

For the early measurements (MJD\,49049 -- MJD\,54226) the apparent disc size is larger than the orbital radius of the NS, with radii ranging between $r \sim 8.6 \times 10^{10}$\,m and $r \sim 9.4 \times 10^{10}$\,m. The two SALT data points are at $r \sim 8.8 \times 10^{10}$\,m, below the orbital radius of the NS. The X-ray behaviour of 1A~1118$-$61 is characterised by long periods of quiescence and rare type \rom{2} outbursts. The system underwent a type \rom{2} outburst in 2009 (peaking at $\sim$MJD\,54845) for which there was no optical coverage. Prior to this, 1A~1118$-$61 displayed one other outburst in January 1992 (MJD 48646) \citep{1994A&A...289..784C} since its discovery in 1974. The very low luminosities of the Type II outbursts ($<10^{37}$\,erg\,s$^{-1}$) in Fig.~\ref{1118rad} suggests that the object is probably less distant from us than the 5\,kpc estimate \citep{JanotPacheco1981} used in this work.

\section{Discussion}
\label{sec:discussion}

\begin{figure*}
\centering
\includegraphics[width=0.45\textwidth, angle=0]{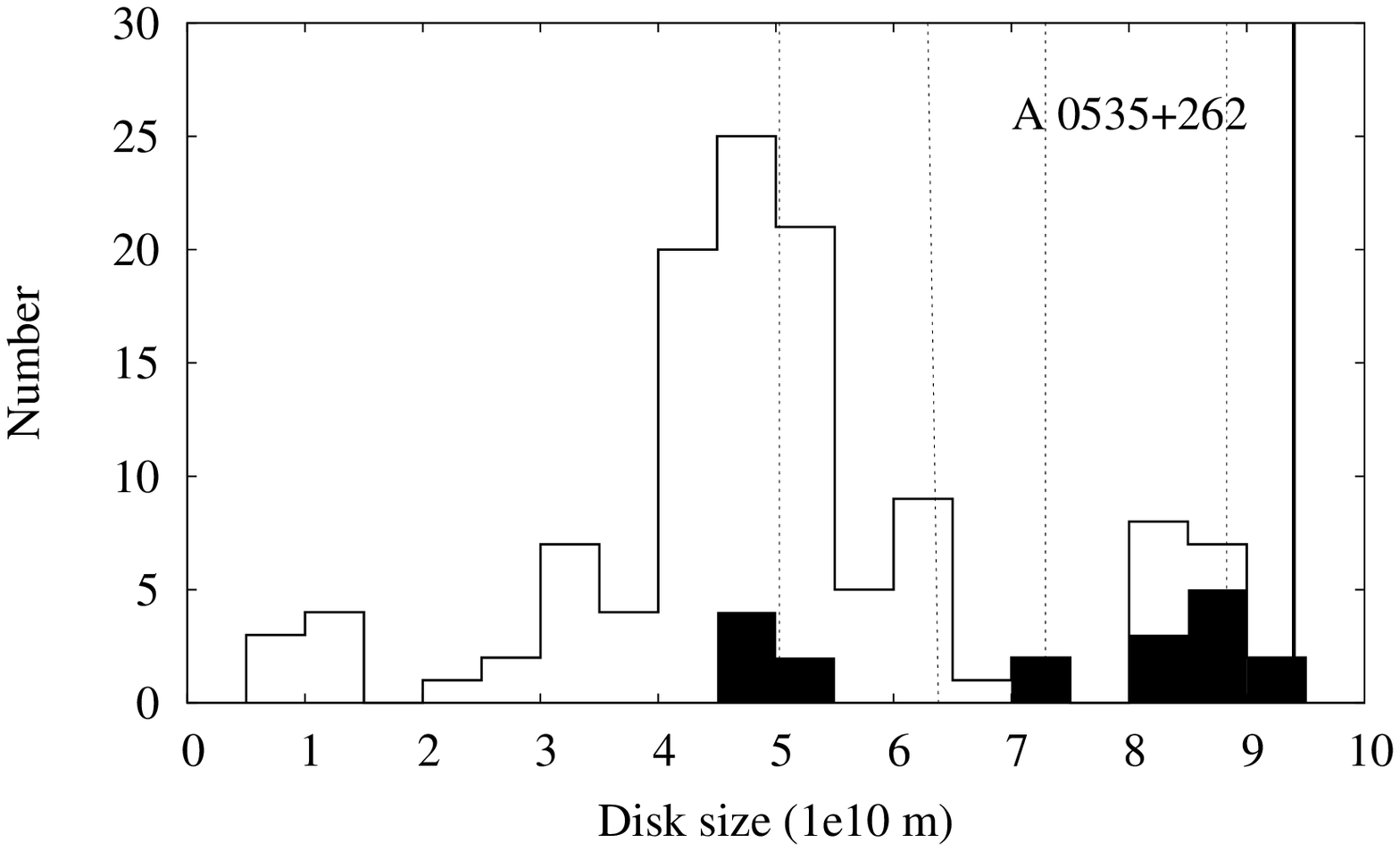}\hspace{0.5cm}\includegraphics[width=0.45\textwidth, angle=0]{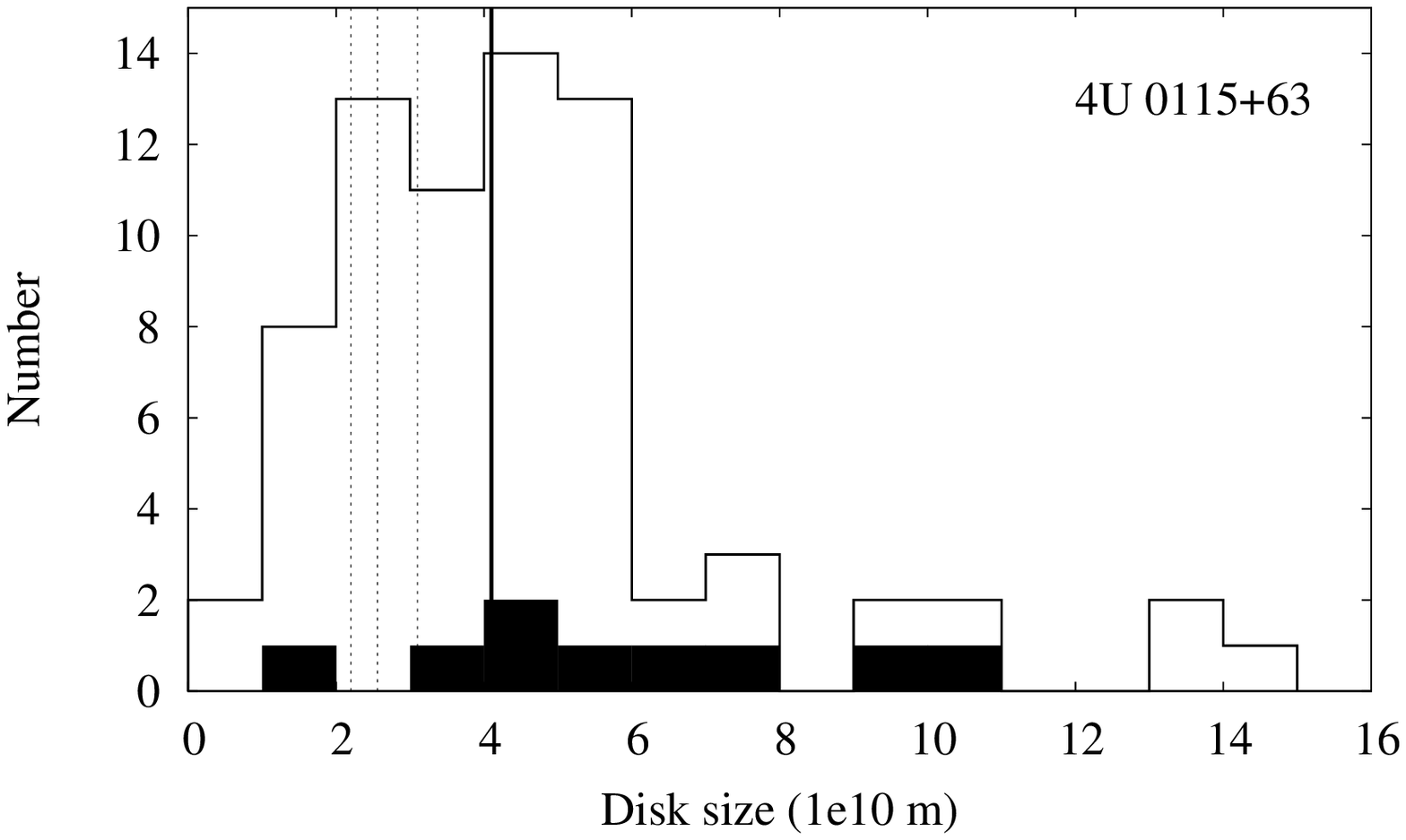}\\
\vspace{0.5cm}
\includegraphics[width=0.45\textwidth, angle =0]{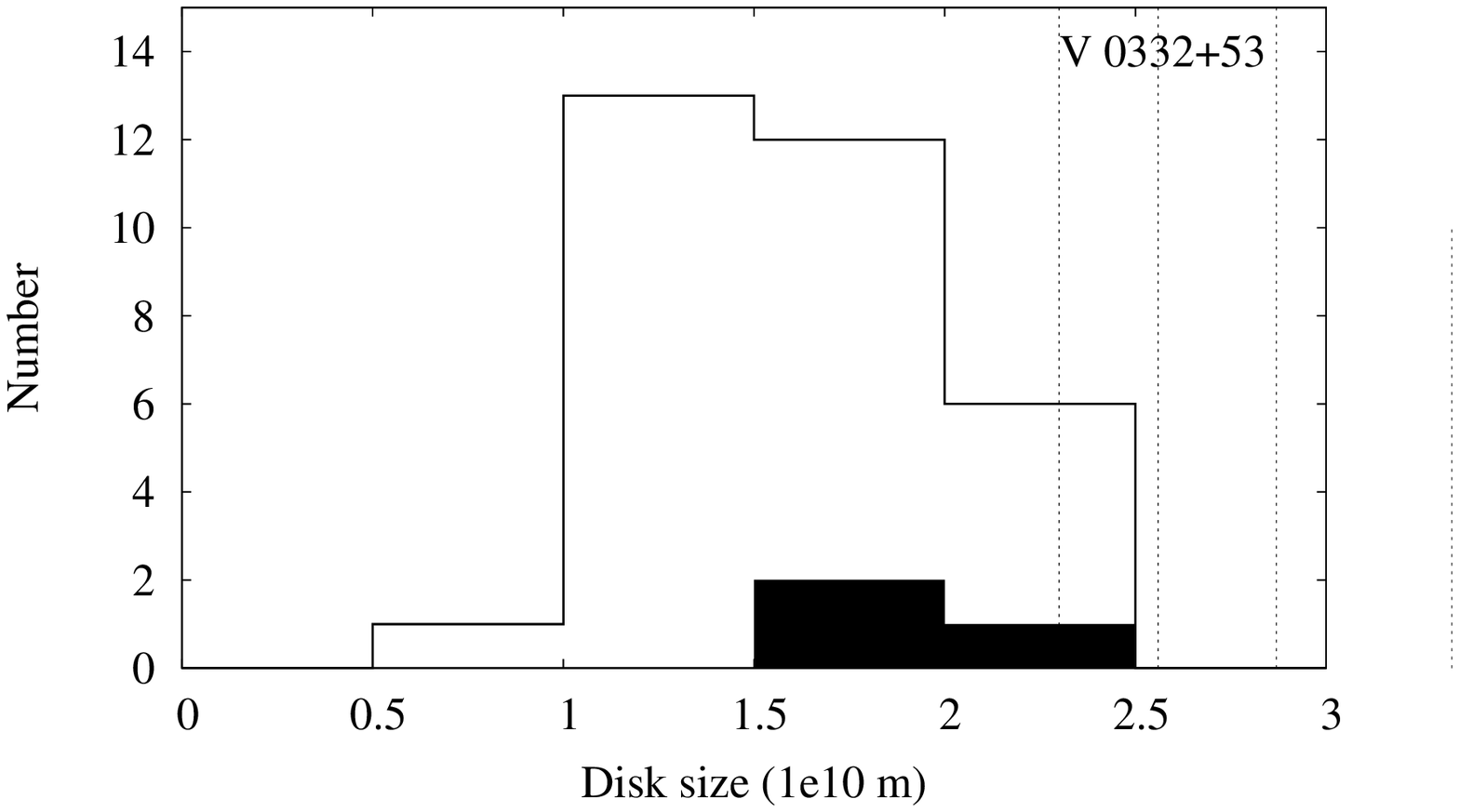}\hspace{0.5cm}\includegraphics[width=0.45\textwidth, angle = 0]{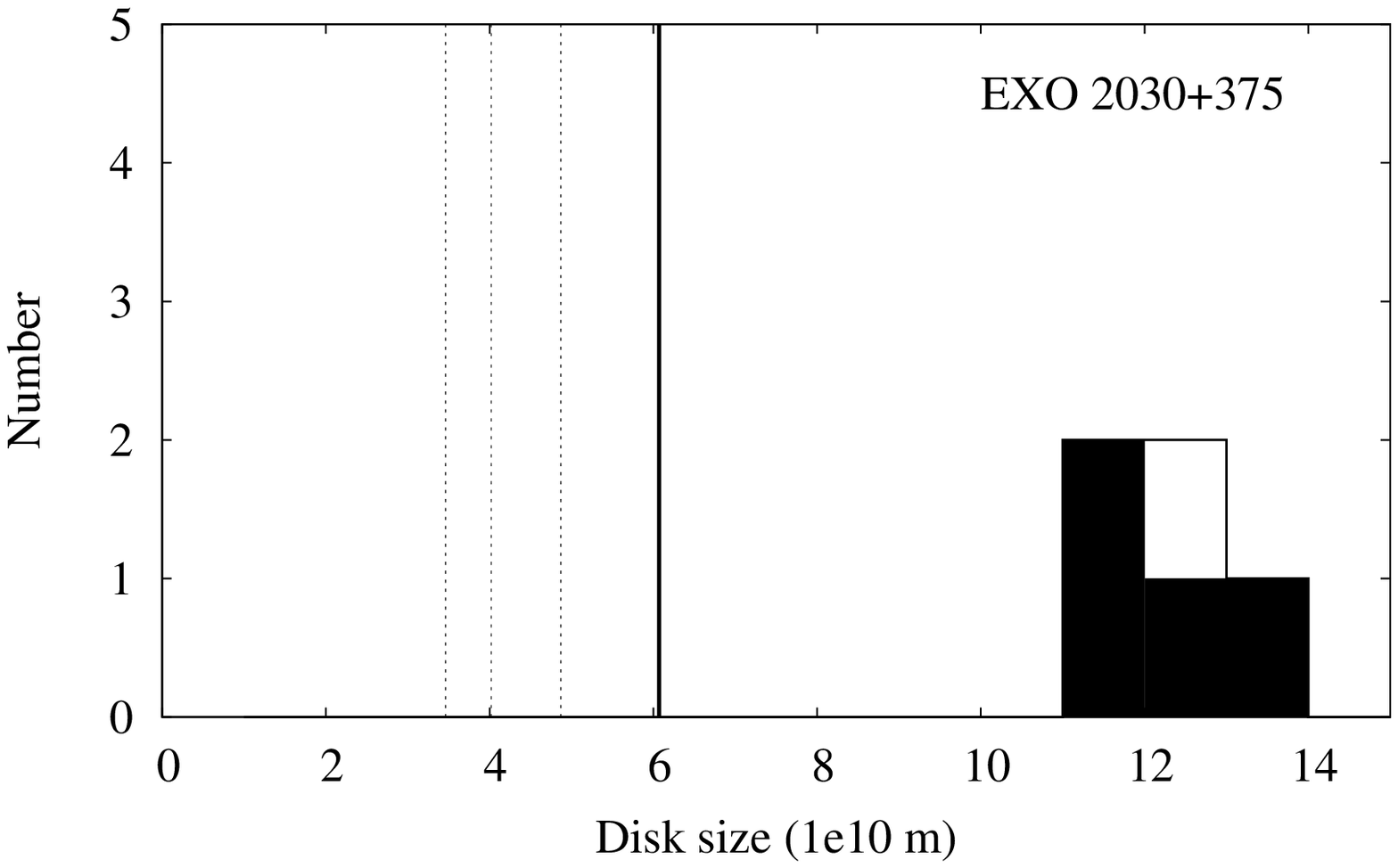}\\
	\caption{The distribution of disc sizes.  Only H$\alpha$ observations during the lifetime of the \emph{RXTE}-ASM and Swift missions are shown. Measurements within 3 days of X-ray activity are highlighted in solid colour. Dotted lines show the 7:1, 5:1, 4:1 and 3:1 resonance radii (from small to large radii) for 1A~0535+262; 5:1, 4:1 and 3:1 resonance radii for 4U~0115+63; 7:1, 6:1 and 5:1 resonance radii for V~0332+53; 5:1, 4:1 and 3:1 resonance radii for EXO~2030+375, while the solid lines show the periastron distance of the neutron star.} 
	\label{fig:histograms}
\end{figure*}

Figure~\ref{fig:histograms} shows the data already presented as histograms of disc size. In these histograms we only plot H$\alpha$ disc size measurements for epochs that are covered by either the \emph{RXTE} All Sky Monitor (ASM) or Swift hard X-ray monitoring \citep{KrimmHollandCorbet2013} data. H$\alpha$ disc size estimates that occur within 3 days of a $5\sigma$ detection with either \emph{RXTE} ASM or Swift-BAT are highlighted with solid colour histograms.  1A~1118$-$61 is not included as there no H$\alpha$ measurements `simultaneous' with X-ray activity. Looking at the evolution of the Be disc radius we are struck by a few interesting results which we discuss below.


\subsection{Type \rom{2} outbursts with small disc sizes}
 In a few instances we see type \rom{2} outbursts occurring around the time when the apparent disc size is smaller than the critical lobe radius\footnote{The critical lobe is the disc radius that reaches the $L_1$ point, i.e. mass transfer can occur when the disc is at, or larger than, the critical lobe.} at periastron passage of the neutron star (R$_\mathrm{crit}$) in 1A~0535+262 and V~0332+53. During the 1994 (MJD\,49500) type \rom{2} outburst the disc size in 1A~0535+262 was below the 4:1 r$_\mathrm{res}$ (4:1 r$_\mathrm{res}$ $\approx$ R$_\mathrm{crit}$), while during the  2011 (MJD\,53500) outburst in this system the disc was below the 5:1  r$_\mathrm{res}$. 
When the November 2004 type \rom{2} outburst had already begun (MJD\,53332) the disc in V0332+53 was still truncated below 5:1 r$_\mathrm{res}$ (3:1 r$_\mathrm{res}$ $\approx$ R$_\mathrm{crit}$), as well as approximately two cycles before the large intensity type I outburst in 2008 (MJD\,55200). The apparent disc radius was still below the 5:1 r$_\mathrm{res}$ at the beginning of the LT monitoring while the system was undergoing a short series of type \rom{1} outbursts.\\
In a very simple model, where the $\Delta V$ and EW(H$\alpha$) trace the disc radius, one would expect the disc to be large enough for material to be accreted to give rise to these outbursts. A number of possible reasons for the low H$\alpha$ emission during the outbursts are discussed below.
\subsubsection{Precessing warped disc}
The H$\alpha$ line profile shape is seen to undergo drastic changes which have been previously studied in 1A~0535+262 \citep{MoritaniNogamiOkazaki2013} and 4U~0115+634 \citep{NegueruelaOkazakiFabregat2001, ReigLarionovNegueruela2007} and were interpreted to have resulted from a precessing warped disc. The low H$\alpha$ emission during type \rom{2} outbursts can be explained by a precessing warped disc, as was suggested to be the scenario for the 2011 outburst in 1A~0535+262 by \citet{MoritaniNogamiOkazaki2013}. Here the changes in the H$\alpha$ line profile were explained as arising from a warped region of the disc not being bright in H$\alpha$ or hidden from the observer's line of sight (see their Fig. 11).

\subsubsection{Elongated Be disc}
Strong disturbances of the Be disc resulting in elongation could explain low H$\alpha$ emission during type II outbursts. An eccentric disc with the elongated region in the direction of the NS at periastron passage could result in type \rom{2} outbursts. The low emission would be a line of sight effect: if the elongated region faces away from the observer, the observer is exposed to a smaller surface area of the disc, hence less H$\alpha$. In recent work, \citet{2014ApJ...790L..34M} use 3D smoothed particle hydrodynamics simulations to suggest that type \rom{2} outbursts result from a highly misaligned decretion disc ($i \geq 60^{o}$) which is eccentric, allowing a large amount of matter to be accreted by the NS at close passage.

\subsection{Type \rom{2} outbursts with very large disc sizes}

On a few occasions we see very large disc sizes for 1A~0535+262, 4U~0115+634 and EXO~2030+375 -- larger than the expected truncation from the viscous decretion disc model. The apparent disc size in these systems sometimes reaches values above the periastron passage of the NS. It would be especially difficult for 4U~0115+634 and EXO~2030+375 to reach such large sizes since they have relatively small orbits and moderate eccentricities.

\citet{MoritaniNogamiOkazaki2013} suggested that a precessing warped disc during the 2009 outburst of 1A~0535+262 resulted in enhanced emission of H$\alpha$ (see their Fig. 11). Another possible cause for the large H$\alpha$ emission, translating to large disc sizes, is an increase in disc base density \citep{SilajJonesTycner2010}. In this case the EW(H$\alpha$) increases without necessarily an increase in disc size.

\subsection{Large disc sizes with no X-ray activity}

The Be disc in 4U~0115+634 appeared to still have sizes above periastron passage at MJD\,53562 with no X-ray activity approximately $\sim$250 days after the end of the type \rom{2} outburst in 2004 ($\sim$10 orbital cycles after the outburst). 1A~1118$-$61, although sparsely sampled in the optical, appears to show disc sizes consistently above the orbital radius of the NS, with unassociated X-ray activity. Again, increase in disc base-density as a result of a large accumulation of matter (not necessarily increase in radial extent) could explain the large emission with no X-ray activity.

\subsection{Evidence for disc truncation}
If truncation plays a dominant role in the size of the Be circumstellar disc, one expects the measurements of disc size to cluster at or below the truncation radii (the dotted lines in Fig. 7). Even in light of the uncertainties in the methodology and the degeneracy between disc size and disc density, there appears to be good evidence for this in both 1A~0535+262 (as discussed in Coe et al. 2006 and Grundstrom et al. 2007) and V~0332+53, and to a lesser extent in 4U~0115+63. The tendency for Be disc sizes in BeXBs to cluster around a specific level, from an observational viewpoint, has been discussed in literature (\citealt{2007ApJ...660.1398G}, \citealt{2013A&A...559A..87Z} and \citealt{2016arXiv160505811Z}). In their Fig. 4, \citet{2016arXiv160505811Z} show that, in the context of the viscous truncated disc model, the discs in gamma-ray binaries generally have sizes larger than the compact object periastron distance, similar to EXO~2030+375 and 1A~1118$-$61 in this study. The truncation of discs below periastron distance in BeXBs may be related to the transient X-ray behaviour in these systems, i.e. on occasions when the disc size reaches values near or greater than periastron distance, X-ray activity is observed. 1A~1118$-$61 and EXO~2030+375 do not show convincing evidence for truncation, despite the well-behaved periodic outbursts exhibited by EXO~2030+375. It is worth noting that the disc size estimates for both these objects are based on single-peaked profiles calibrated through the empirical relationship shown Fig \ref{fit}. Measurements of the disc size based on single peaked profiles have large uncertainties (Section \ref{sec:radius}) and are thus unreliable as indicators of disc truncation.

\section{Conclusion}
We have presented results from long term spectroscopic monitoring of five BeXB systems to study the evolution of Be disc radius and look at its influence on X-ray outbursts, in the context of the viscous decretion disc model. The H$\alpha$ emission line was used to obtain an estimate of the disc size and the main result is that type I outbursts generally occur when the Be disc radius is truncated at radii close to/larger than the critical lobe radius at periastron passage. This provides additional evidence (cf. \citealt{ReigNersesianZezas2016}) for the viscous disc model. Type \rom{2} outbursts, however, are difficult to predict solely based on disc size information, as these are seen to occur when the apparent disc size is large (larger than R$_\mathrm{crit}$), as well as when the disc is smaller than R$_\mathrm{crit}$.

As is becoming clear from other recent analyses of H$\alpha$ data \citep{MoritaniNogamiOkazaki2013} and disc simulations \citep{2014ApJ...790L..34M}, the circumstellar disc behaviour is more complex than just growth, decay and truncation. The large disc sizes estimated on this assumption are unphysical and more complex behaviours such as warping, ellipticity or increases in
the base density must be invoked to explain the observations presented here.

\section*{Acknowledgements}
 IM acknowledges funding from SKA, South Africa. VAM acknowledges funding from the NRF, South Africa and the World Universities Network.

\label{lastpage}

\bibliographystyle{mnras}
\bibliography{references}

\onecolumn

\appendix
\section{Optical spectroscopic data}
\label{append}
The following are the H$_\alpha$ emission line parameters of the BeXBs in study. Columns 1 and 2 list the modified Julian dates and calculated radii, respectively. The measured peak separations and H$\alpha$ equivalent widths are listed in columns 3 and 4, respectively. Column 5 lists the different telescopes used to obtain the spectra: Jacobus Kapetyn Telescope (JKT), Isaac Newton Telescope (INT), Skinakas Telescope (Ski),  South African Astronomical Observatory 1.9-m telescope (SAAO), Russian-Turkish 1.5 m telescope (RT), Liverpool Telescope (LT), Southern African Large Telescope (SALT).\\

\begin{center}
   
\begin{longtable}{| c  c  c  c  c |}
\caption{\textbf{ 1A~0535+262: EW(H$_\alpha$) and $\Delta$V measurements}} \\
\hline
\multicolumn{1}{|c }{{MJD}} & \multicolumn{1}{c}{{radius (m)}} & \multicolumn{1}{c}{{$\Delta$V (km/s)}} & \multicolumn{1}{c}{{EW(H$_{\alpha}$) (\AA) }} & \multicolumn{1}{c|}{{Telescope }}\\ \hline 
\endfirsthead

\multicolumn{3}{c}%
{{\bfseries \tablename\ \thetable{} -- continued from previous page}} \\
\hline
 \multicolumn{1}{|c }{{MJD}} & \multicolumn{1}{c}{{radius (m)}} & \multicolumn{1}{c}{{$\Delta$V (km/s)}} & \multicolumn{1}{c}{{EW(H$_{\alpha}$) (\AA) }} & \multicolumn{1}{c|}{{Telescope }}\\ \hline 
\endhead

\hline \multicolumn{3}{r}{{Continued on next page}} \\ \hline
\endfoot

\hline \hline
\endlastfoot
    
47143  &   5.01 $\times$ 10$^{10}$  & & -11.30 $\pm$ 0.65  &  INT\\
47143  &   4.40 $\times$ 10$^{10}$  & &  -9.38 $\pm$ 0.57  &  INT\\
47153  &   4.65 $\times$ 10$^{10}$  & & -10.20 $\pm$ 0.73  &  INT\\
47160  &   4.06 $\times$ 10$^{10}$  & &  -8.32 $\pm$ 0.89  &  INT\\
47230  &   6.17 $\times$ 10$^{10}$  & & -15.56 $\pm$ 0.91  &  INT\\
47230  &   5.73 $\times$ 10$^{10}$  & & -13.91 $\pm$ 0.23  &  INT\\
48209  &  6.70 $\times$ 10$^{10}$  & 180.2 $\pm$ 7.2  & -11.05 $\pm$ 0.42  &  INT\\
48209  &  6.20 $\times$ 10$^{10}$  & 187.2 $\pm$ 9.4  &  -9.79 $\pm$ 0.30  &  INT\\
48209  &  6.30 $\times$ 10$^{10}$  & 187.1 $\pm$ 7.5  &  -9.81 $\pm$ 0.31  &  INT\\
48252  &   4.90 $\times$ 10$^{10}$  & & -11.02 $\pm$ 0.52  &  INT\\
48252  &  6.00 $\times$ 10$^{10}$  & 191.5 $\pm$ 13.4  & -11.89 $\pm$ 0.96  &  INT\\
48252  &  5.90 $\times$ 10$^{10}$  & 191.9 $\pm$ 7.7  & -11.63 $\pm$ 0.35  &  INT\\
48252  &  6.00 $\times$ 10$^{10}$  & 190.4 $\pm$ 9.5  & -12.15 $\pm$ 0.46  &  INT\\
48283  &  6.30 $\times$ 10$^{10}$  & 186.9 $\pm$ 11.2  & -10.43 $\pm$ 0.63  &  INT\\
48283  &  6.20 $\times$ 10$^{10}$  & 188.4 $\pm$ 7.7  & -10.55 $\pm$ 0.44  &  INT\\
48284  &  6.30 $\times$ 10$^{10}$  & 186.9 $\pm$ 11.2  &  -9.89 $\pm$ 0.76  &  INT\\
48284  &  6.20 $\times$ 10$^{10}$  & 188.5 $\pm$ 13.2  &  -9.88 $\pm$ 0.87  &  INT\\
48362  &  4.30 $\times$ 10$^{10}$  & 225.5 $\pm$ 18.0  &  -7.52 $\pm$ 0.92 &  WHT\\
48362  &  4.30 $\times$ 10$^{10}$  & 225.5 $\pm$ 16.5  &  -7.52 $\pm$ 0.13 &  WHT\\
48496  &  4.60 $\times$ 10$^{10}$  & 218.3 $\pm$ 8.7  &  -9.57 $\pm$ 0.25  &  INT\\
48496  &  4.60 $\times$ 10$^{10}$  & 217.9 $\pm$ 9.8  &  -9.99 $\pm$ 0.70  &  INT\\
48603  &  4.60 $\times$ 10$^{10}$  & 217.1 $\pm$ 10.5  & -12.69 $\pm$ 0.57  &  INT\\
48851  &   3.81 $\times$ 10$^{10}$  & &  -7.55 $\pm$ 0.76  &  INT\\
48852  &  4.50 $\times$ 10$^{10}$  & 219.8 $\pm$ 15.4  &  -8.92 $\pm$ 0.50  &  INT\\
49054  &   6.53 $\times$ 10$^{10}$  & & -16.93 $\pm$ 0.98  &  INT\\
49327  &   6.93 $\times$ 10$^{10}$  & & -18.51 $\pm$ 1.11  &  INT\\
49436  &   4.89 $\times$ 10$^{10}$  & & -10.99 $\pm$ 0.45  &  JKT\\
49436  &   5.57 $\times$ 10$^{10}$  & & -13.35 $\pm$ 0.77  &  JKT\\
49437  &   5.01 $\times$ 10$^{10}$  & & -11.35 $\pm$ 0.86  &  JKT\\
49438  &   5.61 $\times$ 10$^{10}$  & & -13.48 $\pm$ 0.33  &  JKT\\
49611  &  4.00 $\times$ 10$^{10}$  & 233.1 $\pm$ 21.0  & -11.59 $\pm$ 0.21  &  JKT\\
49666  &  4.00 $\times$ 10$^{10}$  & 234.3 $\pm$ 13.5  & -11.66 $\pm$ 0.30  & SAAO\\
49760  &  4.10 $\times$ 10$^{10}$  & 229.8 $\pm$ 13.8  &  -9.77 $\pm$ 0.45  &  JKT\\
49774  &  4.20 $\times$ 10$^{10}$  & 228.4 $\pm$  11.4  & -10.03 $\pm$ 0.50  & SAAO\\
49935  &  4.00 $\times$ 10$^{10}$  & 235.2 $\pm$  10.9  &  -9.68 $\pm$ 0.31  &  JKT\\
50039  &  3.70 $\times$ 10$^{10}$  & 242.6 $\pm$ 9.7  &  -9.39 $\pm$ 0.23  & SAAO\\
50040  &  3.50 $\times$ 10$^{10}$  & 251.7 $\pm$ 10.0  &  -7.76 $\pm$ 1.23  & SAAO\\
50041  &  3.60 $\times$ 10$^{10}$  & 246.8 $\pm$ 11.3  &  -8.50 $\pm$ 0.46  & SAAO\\
50041  &  4.20 $\times$ 10$^{10}$  & 228.7 $\pm$ 13. &  -8.50 $\pm$ 0.98  & SAAO\\
50050  &  3.70 $\times$ 10$^{10}$  & 242.5 $\pm$ 7.1  &  -8.42 $\pm$ 0.16  &  JKT\\
50141  &  3.50 $\times$ 10$^{10}$  & 248.8 $\pm$ 10.5  &  -9.66 $\pm$ 0.25  & SAAO\\
50177  &  3.70 $\times$ 10$^{10}$  & 242.2 $\pm$ 12.2 &  -8.98 $\pm$ 0.51  & SAAO\\
50747  &  3.50 $\times$ 10$^{10}$  & 250.6 $\pm$ 7.9  &  -7.08 $\pm$ 0.69  &  JKT\\
50748  &  3.50 $\times$ 10$^{10}$  & 250.8 $\pm$ 21.2 &  -8.79 $\pm$ 0.44  &  JKT\\
50749  &  3.40 $\times$ 10$^{10}$  & 251.9 $\pm$ 13.4 &  -7.52 $\pm$ 0.76  &  JKT\\
50850  &  3.20 $\times$ 10$^{10}$  & 259.6 $\pm$ 18.8 &  -4.64 $\pm$ 0.92  & SAAO\\
51066  &  7.10 $\times$ 10$^{09}$  & 556.1 $\pm$ 16.0 &  -0.27 $\pm$ 0.21  &  Ski\\
51143  &  9.50 $\times$ 10$^{09}$  & 479.9 $\pm$ 9.0  &  -0.91 $\pm$ 0.20  &  INT\\
51143  &  9.70 $\times$ 10$^{09}$  & 475.0 $\pm$ 13.8  &  -0.96 $\pm$ 0.25  &  INT\\
51171  &  1.00 $\times$ 10$^{10}$  & 464.5 $\pm$ 5.7  &  -0.90 $\pm$ 0.31 &  WHT\\
51213  &  1.10 $\times$ 10$^{10}$  & 446.1 $\pm$ 9.5   &  -0.80 $\pm$ 0.13  &  INT\\
51231  &  1.10 $\times$ 10$^{10}$  & 447.5 $\pm$ 15.3  &  -1.00 $\pm$ 0.22  &  INT\\
51231  &  1.10 $\times$ 10$^{10}$  & 448.4 $\pm$ 7.9   &  -1.06 $\pm$ 0.20  &  INT\\
51241  &  1.10 $\times$ 10$^{10}$  & 442.1 $\pm$ 12.1  &  -0.82 $\pm$ 0.15  &  INT\\
51241  &  1.10 $\times$ 10$^{10}$  & 440.0 $\pm$ 13.2  &  -0.90 $\pm$ 0.34  &  INT\\
51244  &  1.10 $\times$ 10$^{10}$  & 445.8 $\pm$ 18.2  &  -0.85 $\pm$ 0.31  &  INT\\
51471  &  2.30 $\times$ 10$^{10}$  & 311.3 $\pm$ 11.0  &  -3.66 $\pm$ 0.65  &  INT\\
51625  &  2.60 $\times$ 10$^{10}$  & 288.4 $\pm$ 4.8   &  -6.31 $\pm$ 0.91  &  INT\\
51644  &  2.90 $\times$ 10$^{10}$  & 275.2 $\pm$ 19.4   &  -6.48 $\pm$ 0.23  &  INT\\
51771  &  3.30 $\times$ 10$^{10}$  & 257.0 $\pm$ 9.4  &  -6.00 $\pm$ 0.91  &  INT\\
51833  &   4.49 $\times$ 10$^{10}$  & &  -9.57 $\pm$ 0.34  &  Ski\\
51833  &   4.46 $\times$ 10$^{10}$  & &  -9.67 $\pm$ 0.56  &  Ski\\
52164  &   4.57 $\times$ 10$^{10}$  & &  -9.91 $\pm$ 0.11  &  Ski\\
52190  &  4.10 $\times$ 10$^{10}$  & 230.2 $\pm$ 19.6  & -11.14 $\pm$ 0.04  &  Ski\\
52190  &  4.10 $\times$ 10$^{10}$  & 230.3 $\pm$ 17.6  & -11.10 $\pm$ 0.12  &  Ski\\
52528  &   5.82 $\times$ 10$^{10}$  & & -14.27 $\pm$ 0.90 &  Ski\\
52621  &  4.70 $\times$ 10$^{10}$  & 215.6 $\pm$ 11.6  & -14.50 $\pm$ 0.76  &  Ski\\
52918  &   4.93 $\times$ 10$^{10}$  & & -11.12 $\pm$ 0.45  &  Ski\\
52918  &   7.77 $\times$ 10$^{10}$  & & -21.96 $\pm$ 0.32  &  Ski\\
53302  &   6.23 $\times$ 10$^{10}$  & & -15.79 $\pm$ 1.01  &  Ski\\
53448  &   6.11 $\times$ 10$^{10}$  & & -15.33 $\pm$ 0.23  &  Ski\\
53450  &  6.30 $\times$ 10$^{10}$  & 186.3 $\pm$ 7.4  & -15.50 $\pm$ 0.91  &  Ski\\
53598  &   6.20 $\times$ 10$^{10}$  & & -15.69 $\pm$ 0.22  &  Ski\\
53598  &   6.20 $\times$ 10$^{10}$  & & -15.67 $\pm$ 0.46  &  Ski\\
53598  &   6.88 $\times$ 10$^{10}$  & & -18.33 $\pm$ 0.89  &  Ski\\
53599  &   6.19 $\times$ 10$^{10}$  & & -15.63 $\pm$ 0.70  &  Ski\\
53599  &   6.05 $\times$ 10$^{10}$  & & -14.91 $\pm$ 0.57  &  Ski\\
53599  &   6.10 $\times$ 10$^{10}$  & & -15.10 $\pm$ 0.99  &  Ski\\
53599  &   6.20 $\times$ 10$^{10}$  & & -15.62 $\pm$ 0.34  &  Ski\\
54348  &   5.20 $\times$ 10$^{10}$  & & -12.03 $\pm$ 0.78  &  Ski\\
54353  &   5.33 $\times$ 10$^{10}$  & & -12.50 $\pm$ 0.50  &  Ski\\
54711  &   6.19 $\times$ 10$^{10}$  & & -15.63 $\pm$ 0.34  &  Ski\\
55054  &   7.21 $\times$ 10$^{10}$  & & -19.63 $\pm$ 0.56  &  Ski\\
55057  &   7.26 $\times$ 10$^{10}$  & & -19.85 $\pm$ 0.68  &  Ski\\
55104  &   8.14 $\times$ 10$^{10}$  & & -23.55 $\pm$ 0.91  &  Ski\\
55163.13  &   8.72 $\times$ 10$^{10}$  & & -26.09 $\pm$ 0.92  &   LT\\
55167.01  &   8.57 $\times$ 10$^{10}$  & & -25.43 $\pm$ 1.53  &   LT\\
55169.13  &   8.32 $\times$ 10$^{10}$  & & -24.33 $\pm$ 0.43  &   LT\\
55172.13  &   9.29 $\times$ 10$^{10}$  & & -28.70 $\pm$ 0.90  &   LT\\
55175.90  &   8.36 $\times$ 10$^{10}$  & & -24.50 $\pm$ 0.54  &   LT\\
55191.89  &   9.33 $\times$ 10$^{10}$  & & -28.89 $\pm$ 0.55  &   LT\\
55196.89  &   8.64 $\times$ 10$^{10}$  & & -25.76 $\pm$ 0.90  &   LT\\
55199.86  &   8.65 $\times$ 10$^{10}$  & & -25.78 $\pm$ 0.12  &   LT\\
55203.93  &   8.44 $\times$ 10$^{10}$  & & -24.86 $\pm$ 0.45  &   LT\\
55208.91  &   8.43 $\times$ 10$^{10}$  & & -24.82 $\pm$ 0.88  &   LT\\
55211.86  &   8.38 $\times$ 10$^{10}$  & & -24.58 $\pm$ 0.97  &   LT\\
55214.13  &   8.72 $\times$ 10$^{10}$  & & -26.13 $\pm$ 0.43  &   LT\\
55219.02  &   8.80 $\times$ 10$^{10}$  & & -26.46 $\pm$ 0.65  &   LT\\
55220.89  &   8.51 $\times$ 10$^{10}$  & & -25.16 $\pm$ 0.45  &   LT\\
55223.95  &   8.13 $\times$ 10$^{10}$  & & -23.51 $\pm$ 0.71  &   LT\\
55436  &   5.45 $\times$ 10$^{10}$  & & -12.92 $\pm$ 0.43  &  Ski\\
55454  &   5.50 $\times$ 10$^{10}$  & & -13.11 $\pm$ 0.71  &  Ski\\
55495.12  &   4.83 $\times$ 10$^{10}$  & & -10.80 $\pm$ 0.90  &   LT\\
55496.14  &   4.99 $\times$ 10$^{10}$  & & -11.34 $\pm$ 0.42  &   LT\\
55499.23  &   4.99 $\times$ 10$^{10}$  & & -11.22 $\pm$ 0.32  &   LT\\
55502.13  &   4.83 $\times$ 10$^{10}$  & & -10.77 $\pm$ 0.67  &   LT\\
55503  &  8.70 $\times$ 10$^{10}$  & 158.6 $\pm$ 6.3  & -11.13 $\pm$ 0.90  &  Ski\\
55505.13  &   4.80 $\times$ 10$^{10}$  & & -10.70 $\pm$ 0.55  &   LT\\
55508.24  &   4.94 $\times$ 10$^{10}$  & & -11.17 $\pm$ 0.42  &   LT\\
55513.11  &   5.02 $\times$ 10$^{10}$  & & -11.42 $\pm$ 0.55  &   LT\\
55516.02  &   5.14 $\times$ 10$^{10}$  & & -11.82 $\pm$ 0.23  &   LT\\
55517.13  &  5.70 $\times$ 10$^{10}$  & 196.4 $\pm$ 10.1  & -11.58 $\pm$ 0.53  &   LT\\
55520.17  &  6.00 $\times$ 10$^{10}$  & 191.5 $\pm$ 6.7  & -11.82 $\pm$ 0.67  &   LT\\
55523.13  &  6.90 $\times$ 10$^{10}$  & 178.6 $\pm$ 10.4  & -10.94 $\pm$ 0.88  &   LT\\
55545.13  &   4.77 $\times$ 10$^{10}$  & & -10.59 $\pm$ 0.70  &   LT\\
55554.18  &   4.94 $\times$ 10$^{10}$  & & -11.16 $\pm$ 0.50  &   LT\\
55565.13  &   4.83 $\times$ 10$^{10}$  & & -10.78 $\pm$ 0.34  &   LT\\
55572.88  &   4.49 $\times$ 10$^{10}$  & &  -9.65 $\pm$ 0.66 &   LT\\
55574.87  &   4.53 $\times$ 10$^{10}$  & &  -9.79 $\pm$ 0.21  &   LT\\
55578.94  &   4.52 $\times$ 10$^{10}$  & &  -9.75 $\pm$ 0.12  &   LT\\
55580.13  &   4.77 $\times$ 10$^{10}$  & & -10.57 $\pm$ 0.37  &   LT\\
55801  &  4.70 $\times$ 10$^{10}$  & 215.2 $\pm$ 5.9  &  -9.63 $\pm$ 0.31 &  Ski\\
55812  &  4.70 $\times$ 10$^{10}$  & 215.2 $\pm$ 12.1 &  -9.30 $\pm$ 0.36 &  Ski\\
55866  &  4.90 $\times$ 10$^{10}$  & 210.5 $\pm$ 8.1 & -10.85 $\pm$ 0.63  &  Ski\\
55930.92  &  4.20 $\times$ 10$^{10}$  & 227.9 $\pm$ 8.2 &  -9.04 $\pm$ 0.44 &   LT\\
55934.99  &  4.30 $\times$ 10$^{10}$  & 226.8 $\pm$ 9.0 & -10.02 $\pm$ 0.76  &   LT\\
55938.95  &  4.40 $\times$ 10$^{10}$  & 222.1 $\pm$ 5.5  & -10.34 $\pm$ 0.87  &   LT\\
55943.12  &  5.20 $\times$ 10$^{10}$  & 204.9 $\pm$ 9.4 & -10.47 $\pm$ 0.31  &   LT\\
55945.03  &  4.40 $\times$ 10$^{10}$  & 221.9 $\pm$ 11.4  &  -9.88 $\pm$ 00.23 &   LT\\
55946.91  &  4.70 $\times$ 10$^{10}$  & 216.7 $\pm$ 8.7 & -10.41 $\pm$ 1.03  &   LT\\
55966.93  &  4.20 $\times$ 10$^{10}$  & 228.0 $\pm$ 9.1 &  -9.72 $\pm$ 0.26 &   LT\\
55990.90  &  4.20 $\times$ 10$^{10}$  & 227.9 $\pm$ 9.0 &  -8.31 $\pm$ 0.38 &   LT\\
55992.99  &  4.10 $\times$ 10$^{10}$  & 230.6 $\pm$ 9.2  &  -7.61 $\pm$ 0.16 &   LT\\
55994.92  &  4.20 $\times$ 10$^{10}$  & 229.5 $\pm$ 11.5  &  -7.81 $\pm$ 0.25 &   LT\\
56173.16  &  4.30 $\times$ 10$^{10}$  & 224.9 $\pm$ 9.0  &  -8.87 $\pm$ 0.51 &   LT\\
56177.16  &  5.40 $\times$ 10$^{10}$  & 201.9 $\pm$ 12.1  &  -8.81 $\pm$ 0.79 &   LT\\
56199.24  &  4.80 $\times$ 10$^{10}$  & 212.9 $\pm$ 8.5  &  -9.00 $\pm$ 0.44 &   LT\\
56225.11  &  5.00 $\times$ 10$^{10}$  & 209.6 $\pm$ 10.5  &  -9.03 $\pm$ 0.76 &   LT\\
56227.02  &  5.10 $\times$ 10$^{10}$  & 208.1 $\pm$ 14.6  &  -9.38 $\pm$ 0.92 &   LT\\
56254.94  &  4.30 $\times$ 10$^{10}$  & 224.9 $\pm$ 9.0  &  -9.66 $\pm$ 0.21 &   LT\\
56269.05  &  4.80 $\times$ 10$^{10}$  & 214.6 $\pm$ 8.6  & -10.58 $\pm$ 0.20  &   LT\\
56285.04  &  4.90 $\times$ 10$^{10}$  & 211.9 $\pm$ 12.7  & -11.46 $\pm$ 0.57  &   LT\\
56290.92  &   4.71 $\times$ 10$^{10}$  & & -10.38 $\pm$ 0.75  &   LT\\
56292.96  &   4.65 $\times$ 10$^{10}$  & & -10.18 $\pm$ 0.73  &   LT\\
56295.11  &  5.10 $\times$ 10$^{10}$  & 206.7 $\pm$ 8.3  & -11.84 $\pm$ 0.98  &   LT\\
56301  &  5.20 $\times$ 10$^{10}$  & 205.2 $\pm$ 9.0  & -11.71 $\pm$ 0.67  &   LT\\
56308.99  &  5.80 $\times$ 10$^{10}$  & 193.9 $\pm$ 7.8  & -12.78 $\pm$ 0.50  &   LT\\
56322.97  &  6.10 $\times$ 10$^{10}$  & 189.1 $\pm$ 13.2  & -12.04 $\pm$ 0.27  &   LT\\
56332.90  &  5.20 $\times$ 10$^{10}$  & 204.3 $\pm$ 12.2  & -12.75 $\pm$ 0.54  &   LT\\
56332.99  &  5.00 $\times$ 10$^{10}$  & 208.9 $\pm$ 10.5  & -12.65 $\pm$ 0.43  &   LT\\
56334.90  &  5.10 $\times$ 10$^{10}$  & 207.8 $\pm$ 8.3  & -12.91 $\pm$ 0.55  &   LT\\
56337.05  &  4.60 $\times$ 10$^{10}$  & 217.8 $\pm$ 10.9  & -12.64 $\pm$ 0.43  &   LT\\
56340.96  &  4.20 $\times$ 10$^{10}$  & 227.3 $\pm$ 9.1 & -13.06 $\pm$ 1.12  &   LT\\
56346.95  &  4.10 $\times$ 10$^{10}$  & 231.3 $\pm$ 12.7  & -12.55 $\pm$ 0.80  &   LT\\
56348.89  &  4.10 $\times$ 10$^{10}$  & 230.1 $\pm$ 13.8  & -13.32 $\pm$ 0.51  &   LT\\
56583  &   5.98 $\times$ 10$^{10}$  & & -14.83 $\pm$ 0.67  &  Ski\\
56583.04  &   5.40 $\times$ 10$^{10}$  & & -12.74 $\pm$ 0.78  &   LT\\
56585.05  & 5.28 $\times$ 10$^{10}$  & 203.7 $\pm$ 9.2  & -14.81 $\pm$ 0.98  &   LT\\
56611.17  & 5.28 $\times$ 10$^{10}$  & 203.5 $\pm$ 10.7  & -14.31 $\pm$ 0.54  &   LT\\
56647.07  & 5.33 $\times$ 10$^{10}$  & 202.7 $\pm$ 8.1  & -14.30 $\pm$ 0.34  &   LT\\
56649.13  & 5.30 $\times$ 10$^{10}$  & 203.3 $\pm$ 8.1  & -14.29 $\pm$ 0.68  &   LT\\
56913.12  &   4.68 $\times$ 10$^{10}$  & & -10.30 $\pm$ 0.90  &   LT\\
56925.11  &   4.42 $\times$ 10$^{10}$  & &  -9.44 $\pm$ 0.21  &   LT\\
57020.96  & 6.14 $\times$ 10$^{10}$  & 188.9 $\pm$ 7.6  &  -8.34 $\pm$ 0.33  &   LT\\
57022.99  & 5.14 $\times$ 10$^{10}$  & 206.4 $\pm$ 8.2  &  -9.53 $\pm$ 0.44  &   LT\\
57057.09  & 5.32 $\times$ 10$^{10}$  & 202.8 $\pm$ 4.1  &  -8.02 $\pm$ 0.60  &   LT\\
57059.06  & 5.20 $\times$ 10$^{10}$  & 205.3 $\pm$ 10.3 &  -8.58 $\pm$ 0.27  &   LT\\
57375.22  &   4.30 $\times$ 10$^{10}$  & &  -9.06 $\pm$ 0.45  &   LT\\
57376.88  & 5.03 $\times$ 10$^{10}$  & 208.6 $\pm$ 13.6  &  -9.27 $\pm$ 0.98  &   LT\\
57384.94  & 5.41 $\times$ 10$^{10}$  & 201.1 $\pm$ 13.3  &  -9.48 $\pm$ 0.57  &   LT\\   \end{longtable}
   \end{center}

 \begin{center}
   
   \begin{longtable}{| c  c  c  c  c |}
   \caption{\textbf{4U~0115+63: EW(H$_\alpha$) and $\Delta$V measurements}} \\
   \hline
   \multicolumn{1}{|c }{{MJD}} & \multicolumn{1}{c}{{radius (m)}} & \multicolumn{1}{c}{{$\Delta$V (km/s)}} & \multicolumn{1}{c}{{EW(H$_{\alpha}$) (\AA) }} & \multicolumn{1}{c|}{{Telescope }}\\ \hline 
\endfirsthead

\multicolumn{3}{c}%
{{\bfseries \tablename\ \thetable{} -- continued from previous page}} \\
\hline
 \multicolumn{1}{|c }{{MJD}} & \multicolumn{1}{c}{{radius (m)}} & \multicolumn{1}{c}{{$\Delta$V (km/s)}} & \multicolumn{1}{c}{{EW(H$_{\alpha}$) (\AA) }} & \multicolumn{1}{c|}{{Telescope }}\\ \hline 
\endhead

\hline \multicolumn{3}{r}{{Continued on next page}} \\ \hline
\endfoot

\hline \hline
\endlastfoot

    \hline
47936 & 4.32 $\times$ 10$^{10}$ &   & -9.36 $\pm$ 0.97& WHT\\
47936 & 4.93 $\times$ 10$^{10}$ &  & -11.40 $\pm$ 1.19& WHT\\
48252 & 2.40 $\times$ 10$^{10}$ &    430.0 $\pm$ 14.8  & -2.94 $\pm$ 0.89& INT\\
48252 & 2.78 $\times$ 10$^{10}$ &    400.0 $\pm$ 17.2  & -2.89 $\pm$ 0.51& INT\\
48252 & 2.52 $\times$ 10$^{10}$ &    420.0 $\pm$ 16.0  & -5.19 $\pm$ 0.37& INT\\
48252 & 2.78 $\times$ 10$^{10}$ &    400.0 $\pm$ 16.9  & -4.81 $\pm$ 0.51& INT\\
48283 & 5.20 $\times$ 10$^{10}$ &  & -12.33 $\pm$ 0.67& INT\\
48283 & 5.35 $\times$ 10$^{10}$ &  & -12.89 $\pm$ 0.90& INT\\
48283 & 5.02 $\times$ 10$^{10}$ &  & -11.72 $\pm$ 0.83& INT\\
48496 & 2.01 $\times$ 10$^{10}$ &    470.0 $\pm$ 7.9  & -5.66 $\pm$ 0.48& INT\\
48604 & 1.85 $\times$ 10$^{10}$ &    490.0 $\pm$ 15.2  & -6.71 $\pm$ 0.26& INT\\
48604 & 1.93 $\times$ 10$^{10}$ &    480.0 $\pm$ 20.6  & -5.83 $\pm$ 0.28& INT\\
48839 & 2.52 $\times$ 10$^{10}$ &    420.0 $\pm$ 18.0  & -4.37 $\pm$ 0.51& INT\\
48839 & 2.30 $\times$ 10$^{10}$ &    440.0 $\pm$ 22.6  & -4.33 $\pm$ 0.79& INT\\
49005 & 2.30 $\times$ 10$^{10}$ &    440.0 $\pm$ 19.2  & -7.77 $\pm$ 0.44& INT\\
49338 & 2.64 $\times$ 10$^{10}$ &    410.0 $\pm$ 16.5  & -6.83 $\pm$ 0.76& WHT\\
49338 & 3.08 $\times$ 10$^{10}$ &    380.0 $\pm$ 18.2  & -6.69 $\pm$ 0.92& WHT\\
49338 & 3.08 $\times$ 10$^{10}$ &    380.0 $\pm$ 15.1  & -7.54 $\pm$ 0.57& WHT\\
49340 & 2.78 $\times$ 10$^{10}$ &    400.0 $\pm$ 17.6  & -6.25 $\pm$ 0.34& WHT\\
49340 & 2.92 $\times$ 10$^{10}$ &    390.0 $\pm$ 17.3  & -6.13 $\pm$ 0.59& WHT\\
49726 & 4.39 $\times$ 10$^{10}$ &  & -9.57  $\pm$ 0.54& INT\\
49901 & 2.10 $\times$ 10$^{10}$ &    460.0 $\pm$ 7.0  & -4.24 $\pm$ 0.21& INT\\
49901 & 2.20 $\times$ 10$^{10}$ &    450.0 $\pm$ 7.1  & -5.45 $\pm$ 0.65& INT\\
49972 & 2.52 $\times$ 10$^{10}$ &    420.0 $\pm$ 8.2  & -5.28 $\pm$ 0.23& INT\\
49972 & 2.64 $\times$ 10$^{10}$ &    410.0 $\pm$ 10.7  & -5.34 $\pm$ 0.16& INT\\
50050 & 5.41 $\times$ 10$^{10}$ & & -13.10 $\pm$ 1.00 & JKT\\
50094 & 7.72 $\times$ 10$^{10}$ &    240.0 $\pm$ 9.6 & -12.81 $\pm$ 0.82& INT\\
50094 & 7.72 $\times$ 10$^{10}$ &    240.0 $\pm$ 13.9 & -12.41 $\pm$ 0.90& INT\\
50094 & 7.72 $\times$ 10$^{10}$ &    240.0 $\pm$ 8.6 & -11.86 $\pm$ 0.45& INT\\
50113 & 4.34 $\times$ 10$^{10}$ &    320.0 $\pm$ 8.3  & -7.54 $\pm$ 0.30& INT\\
50113 & 4.34 $\times$ 10$^{10}$ &    320.0 $\pm$ 11.6  & -8.61 $\pm$ 0.25& INT\\
50254 & 3.85 $\times$ 10$^{10}$ &    340.0 $\pm$ 11.7  & -7.84 $\pm$ 0.94& INT\\
50273 & 2.52 $\times$ 10$^{10}$ &    420.0 $\pm$ 16.1  & -6.61 $\pm$ 0.43& INT\\
50273 & 2.52 $\times$ 10$^{10}$ &    420.0 $\pm$ 13.4  & -6.65 $\pm$ 0.55& INT\\
50480 & 2.01 $\times$ 10$^{10}$ &    470.0 $\pm$ 11.0  & -4.34 $\pm$ 0.23& INT\\
51026 & 3.63 $\times$ 10$^{10}$ &    350.0 $\pm$ 10.3  & -7.21 $\pm$ 1.22& INT\\
51066 & 3.85 $\times$ 10$^{10}$ &    340.0 $\pm$ 13.6  & -7.43 $\pm$ 0.80& INT\\
51084 & 4.63 $\times$ 10$^{10}$ &    310.0 $\pm$ 12.4  & -7.98 $\pm$ 0.51& INT\\
51175 & 4.89 $\times$ 10$^{10}$ & & -11.27 $\pm$ 0.57& INT\\
51385 & 5.51 $\times$ 10$^{10}$ & & -13.47 $\pm$ 0.57& Ski\\
51385 & 1.11 $\times$ 10$^{11}$ &    200.0 $\pm$ 8.7   & -9.18 $\pm$ 0.20& Ski\\
51743 & 1.54 $\times$ 10$^{11}$ &    170.0 $\pm$ 10.2 & -12.01 $\pm$ 0.15& Ski\\
51743 & 1.37 $\times$ 10$^{11}$ &    180.0 $\pm$ 12.6 & -12.69 $\pm$ 0.34& Ski\\
51743 & 1.54 $\times$ 10$^{11}$ &    170.0 $\pm$ 11.1 & -12.45 $\pm$ 0.31& Ski\\
51822 & 9.18 $\times$ 10$^{10}$ &    220.0 $\pm$ 19.4 & -20.84 $\pm$ 0.65& Ski\\
51833 & 5.67 $\times$ 10$^{10}$ &    280.0 $\pm$ 9.7 & -20.04 $\pm$ 0.91& Ski\\
52190 & 6.95 $\times$ 10$^{9}$  &  800.0 $\pm$ 20.1  & -2.48 $\pm$ 0.23& Ski\\
52191 & 1.05 $\times$ 10$^{10}$ &    650.0 $\pm$ 14.5  & -2.47 $\pm$ 0.91& Ski\\
52918 & 2.20 $\times$ 10$^{10}$ &    450.0 $\pm$ 10.0 & -3.42 $\pm$ 0.34& Ski\\
53180 & 1.37 $\times$ 10$^{11}$ &    180.0 $\pm$ 12.0 & -9.03 $\pm$ 0.65& Ski\\
53193 & 1.37 $\times$ 10$^{11}$ &    180.0 $\pm$ 11.2 & -9.29 $\pm$ 0.37& Ski\\
53243 & 1.01 $\times$ 10$^{11}$ &    210.0 $\pm$ 17.6 & -9.26 $\pm$ 0.56& Ski\\
53252 & 6.10 $\times$ 10$^{10}$ &    270.0 $\pm$ 9.8 & -9.51 $\pm$ 0.59& Ski\\
53260 & 4.60 $\times$ 10$^{10}$ & & -10.13 $\pm$ 0.14& Ski\\
53261 & 5.29 $\times$ 10$^{10}$ &    290.0 $\pm$ 25.2  & -9.27 $\pm$ 0.97& Ski\\
53303 & 5.84 $\times$ 10$^{10}$ & & -14.68 $\pm$ 0.45& Ski\\
53544 & 2.78 $\times$ 10$^{10}$ &    400.0 $\pm$ 17.6  & -9.51 $\pm$ 0.97& Ski\\
53562 & 3.85 $\times$ 10$^{10}$ &    340.0 $\pm$ 9.8 & -12.34 $\pm$ 0.92& Ski\\
53580 & 5.67 $\times$ 10$^{10}$ &    280.0 $\pm$ 11.8 & -17.57 $\pm$ 0.23& Ski\\
53599 & 6.58 $\times$ 10$^{10}$ &    260.0 $\pm$ 13.6 & -16.44 $\pm$ 0.44& Ski\\
53600 & 7.11 $\times$ 10$^{10}$ &    250.0 $\pm$ 14.9 & -17.88 $\pm$ 0.64& Ski\\
53633 & 4.94 $\times$ 10$^{10}$ &    300.0 $\pm$ 13.5 & -15.11 $\pm$ 0.89& Ski\\
53669 & 5.67 $\times$ 10$^{10}$ &    280.0 $\pm$ 17.2  & -6.69 $\pm$ 0.55& Ski\\
54010 & 2.30 $\times$ 10$^{10}$ &    440.0 $\pm$ 28.8  & -4.88 $\pm$ 0.46& Ski\\
54348 & 7.11 $\times$ 10$^{10}$ &    250.0 $\pm$ 12.5  & -9.88 $\pm$ 0.76& Ski\\
54375 & 4.94 $\times$ 10$^{10}$ &    300.0 $\pm$ 13.1  & -9.63 $\pm$ 0.90& Ski\\
54377 & 3.85 $\times$ 10$^{10}$ &    340.0 $\pm$ 13.7  & -8.76 $\pm$ 0.92& Ski\\
54641 & 3.25 $\times$ 10$^{10}$ &    370.0 $\pm$ 20.4 & -11.85 $\pm$ 0.94& Ski\\
54643 & 3.85 $\times$ 10$^{10}$ &    340.0 $\pm$ 13.9 & -12.69 $\pm$ 0.93& Ski\\
54662 & 5.67 $\times$ 10$^{10}$ & & -14.07 $\pm$ 0.57& Ski\\
54686 & 2.40 $\times$ 10$^{10}$ &    430.0 $\pm$ 17.6 & -12.92 $\pm$ 0.22& Ski\\
55041 & 8.57 $\times$ 10$^{9}$  &  720.0 $\pm$ 7.9  & -1.46 $\pm$ 0.47& Ski\\
55220.86 & 1.37 $\times$ 10$^{10}$ &    570.0 $\pm$ 16.8  & -4.25 $\pm$ 0.53& LT\\
55347.21 & 1.93 $\times$ 10$^{10}$ &    480.0 $\pm$ 16.9  & -4.54 $\pm$ 0.91& LT\\
55361.13 & 2.64 $\times$ 10$^{10}$ &    410.0 $\pm$ 18.0  & -5.79 $\pm$ 0.87& LT\\
55361.14 & 2.20 $\times$ 10$^{10}$ &    450.0 $\pm$ 15.9  & -7.11 $\pm$ 0.23& LT\\
55361.15 & 3.08 $\times$ 10$^{10}$ &    380.0 $\pm$ 25.5  & -8.15 $\pm$ 0.65& LT\\
55375.13 & 2.30 $\times$ 10$^{10}$ &    440.0 $\pm$ 18.0  & -5.59 $\pm$ 0.89& LT\\
55375.14 & 2.40 $\times$ 10$^{10}$ &    430.0 $\pm$ 18.8  & -6.76 $\pm$ 0.91& LT\\
55375.15 & 3.25 $\times$ 10$^{10}$ &    370.0 $\pm$ 8.1  & -4.82 $\pm$ 0.67& LT\\
55403.21 & 4.43 $\times$ 10$^{10}$ &   & -9.70 $\pm$ 0.47& LT\\
55403.21 & 4.28 $\times$ 10$^{10}$ &   & -9.23 $\pm$ 0.59& LT\\
55403.22 & 4.11 $\times$ 10$^{10}$ &   & -8.70 $\pm$ 0.34& LT\\
55410 & 4.71 $\times$ 10$^{10}$ &  & -10.63 $\pm$ 0.51& LT\\
55435 & 4.63 $\times$ 10$^{10}$ &    310.0 $\pm$ 12.4  & -6.61 $\pm$ 0.67& Ski\\
55437 & 4.08 $\times$ 10$^{10}$ &    330.0 $\pm$ 16.5  & -7.14 $\pm$ 0.50& Ski\\
55453 & 3.85 $\times$ 10$^{10}$ &    340.0 $\pm$ 4.1 & -7.24 $\pm$ 0.76& Ski\\
55469 & 3.64 $\times$ 10$^{10}$ &  & -7.22 $\pm$ 0.33& Ski\\
55731.18 & 1.11 $\times$ 10$^{11}$ &    200.0 $\pm$ 9.8  & -9.04 $\pm$ 0.54& LT\\
55738.20 & 4.64 $\times$ 10$^{10}$ & & -10.40 $\pm$ 0.65& LT\\
55751.17 & 3.08 $\times$ 10$^{10}$ &    380.0 $\pm$ 28.8  & -9.12 $\pm$ 0.23& LT\\
55758.15 & 1.64 $\times$ 10$^{10}$ &    520.0 $\pm$ 12.5  & -8.21 $\pm$ 0.56& LT\\
55793 & 5.29 $\times$ 10$^{10}$ &    290.0 $\pm$ 13.1 & -13.13 $\pm$ 1.05& Ski\\
55801 & 5.67 $\times$ 10$^{10}$ &    280.0 $\pm$ 13.7 & -13.18 $\pm$ 1.06& Ski\\
55812 & 2.52 $\times$ 10$^{10}$ &    420.0 $\pm$ 11.6 & -14.60 $\pm$ 1.17& Ski\\
55812 & 2.52 $\times$ 10$^{10}$ &    420.0 $\pm$ 11.2 & -19.89 $\pm$ 1.39& Ski\\
55834.99 & 5.75 $\times$ 10$^{10}$ &  & -14.34 $\pm$ 1.29& LT\\
55841.99 & 5.97 $\times$ 10$^{10}$ &  & -15.17 $\pm$ 1.27& LT\\
55897.84 & 4.63 $\times$ 10$^{10}$ &  & -10.37 $\pm$ 0.90& LT\\
55904.90 & 3.39 $\times$ 10$^{10}$ &   & -6.50 $\pm$ 0.52& LT\\
55925.84 & 4.62 $\times$ 10$^{10}$ &  & -10.36 $\pm$ 0.89& LT\\
55941.98 & 5.23 $\times$ 10$^{10}$ &  & -12.44 $\pm$ 0.74& LT\\
55946.90 & 4.73 $\times$ 10$^{10}$ &  & -10.73 $\pm$ 0.97& LT\\
56084 & 2.20 $\times$ 10$^{10}$ &    450.0 $\pm$ 18.0  & -3.87 $\pm$ 0.31& Ski\\
56163 & 1.05 $\times$ 10$^{10}$ &    650.0 $\pm$ 19.5  & -2.22 $\pm$ 0.18& Ski\\
56183 & 1.09 $\times$ 10$^{10}$ &    640.0 $\pm$ 18.3  & -1.58 $\pm$ 0.16& Ski\\
56466.13 & 2.99 $\times$ 10$^{10}$ &  & -5.40 $\pm$ 0.43& LT\\
56472.13 & 2.20 $\times$ 10$^{10}$ &    450.0 $\pm$ 18.0  & -5.50 $\pm$ 0.51& LT\\
56494.06 & 3.00 $\times$ 10$^{10}$ &  & -5.42 $\pm$ 0.43& LT\\
56503 & 2.20 $\times$ 10$^{10}$ &    450.0 $\pm$ 28.4  & -7.55 $\pm$ 0.76& Ski\\
56534 & 2.01 $\times$ 10$^{10}$ &    470.0 $\pm$ 4.7  & -5.86 $\pm$ 0.87& Ski\\
  \end{longtable}
  \end{center}
  
\newpage
   \begin{center}
   
   \begin{longtable}{| c  c  c  c  c |}
   \caption{\textbf{V~0332+53: EW(H$_\alpha$) and $\Delta$V measurements}} \\
   \hline
   \multicolumn{1}{|c }{{MJD}} & \multicolumn{1}{c}{{radius (m)}} & \multicolumn{1}{c}{{$\Delta$V (km/s)}} & \multicolumn{1}{c}{{EW(H$_{\alpha}$) (\AA) }} & \multicolumn{1}{c|}{{Telescope }}\\ \hline 
\endfirsthead

\multicolumn{3}{c}%
{{\bfseries \tablename\ \thetable{} -- continued from previous page}} \\
\hline
 \multicolumn{1}{|c }{{MJD}} & \multicolumn{1}{c}{{radius (m)}} & \multicolumn{1}{c}{{$\Delta$V (km/s)}} & \multicolumn{1}{c}{{EW(H$_{\alpha}$) (\AA) }} & \multicolumn{1}{c|}{{Telescope }}\\ \hline 
\endhead

\hline \multicolumn{3}{r}{{Continued on next page}} \\ \hline
\endfoot

\hline \hline
\endlastfoot

    \hline

47919 &1.70 $\times$ 10$^{10}$ &  & -6.86  $\pm$ 0.58& INT\\
48136 &1.51 $\times$ 10$^{10}$ &    150.0 $\pm$ 3.7  & -7.63  $\pm$ 0.53& INT\\
48136 &1.73 $\times$ 10$^{10}$ &    140.0 $\pm$ 10.9  & -7.48  $\pm$ 0.67& INT\\
48136 &1.73 $\times$ 10$^{10}$ &    140.0 $\pm$ 8.8  & -6.69  $\pm$ 0.88& INT\\
48208 &1.73 $\times$ 10$^{10}$ &    140.0 $\pm$ 10.4  & -6.57  $\pm$ 0.25& INT\\
48209 &1.80 $\times$ 10$^{10}$ &   & -7.24  $\pm$ 0.34& INT\\
48209 &2.00 $\times$ 10$^{10}$ &   & -8.82  $\pm$ 0.21& INT\\
48209 &1.70 $\times$ 10$^{10}$ &   & -7  $\pm$ 0.89& INT\\
48209 &1.70 $\times$ 10$^{10}$ &   & -7.11  $\pm$ 0.80& INT\\
48252 &1.80 $\times$ 10$^{10}$ &   & -7.73  $\pm$ 0.33& INT\\
48252 &2.00 $\times$ 10$^{10}$ &   & -8.70  $\pm$ 0.79& INT\\
48252 &2.40 $\times$ 10$^{10}$ &  & -11.44  $\pm$ 0.93& INT\\
48283 &1.90 $\times$ 10$^{10}$ &   & -7.79  $\pm$ 0.45& INT\\
48283 &1.80 $\times$ 10$^{10}$ &   & -7.52  $\pm$ 0.90& INT\\
48496 &1.51 $\times$ 10$^{10}$ &    150.0 $\pm$ 3.5  & -5.45  $\pm$ 0.50& INT\\
48603 &1.73 $\times$ 10$^{10}$ &    140.0 $\pm$ 7.0  & -8.04  $\pm$ 0.67& INT\\
48604 &1.33 $\times$ 10$^{10}$ &    160.0 $\pm$ 12.8  & -5.24  $\pm$ 0.95& INT\\
48852 &1.73 $\times$ 10$^{10}$ &    140.0 $\pm$ 9.1  & -5.01  $\pm$ 0.23& INT\\
48939 &1.60 $\times$ 10$^{10}$ &  & -6.54  $\pm$ 0.76& INT\\
48939 &1.18 $\times$ 10$^{10}$ &    170.0 $\pm$ 6.8  & -4.23  $\pm$ 0.91& INT\\
48939 &1.73 $\times$ 10$^{10}$ &    140.0 $\pm$ 7.4  & -5.11  $\pm$ 0.34& INT\\
49326 &1.73 $\times$ 10$^{10}$ &    140.0 $\pm$ 7.1 & -10.06  $\pm$ 0.12& INT\\
51744 &1.73 $\times$ 10$^{10}$ &    140.0 $\pm$ 12.0  & -5.33 $\pm$ 0.81& Ski\\
51833 &2.01 $\times$ 10$^{10}$ &    130.0 $\pm$ 10.4  & -5.14 $\pm$ 0.57& Ski\\
51833 &2.01 $\times$ 10$^{10}$ &    130.0 $\pm$ 4.2  & -5.08 $\pm$ 0.57& Ski\\
52128 &1.30 $\times$ 10$^{10}$ &   & -4.65 $\pm$ 0.50& Ski\\
52128 &1.20 $\times$ 10$^{10}$ &   & -3.87 $\pm$ 0.33& Ski\\
52165 &1.50 $\times$ 10$^{10}$ &   & -5.90 $\pm$ 0.91& Ski\\
52918 &1.20 $\times$ 10$^{10}$ &   & -4.12 $\pm$ 0.73& Ski\\
53303 &1.90 $\times$ 10$^{10}$ &   & -8.16 $\pm$ 0.86& Ski\\
53580 &1.40 $\times$ 10$^{10}$ &   & -5.19 $\pm$ 0.23& Ski\\
53598 &1.60 $\times$ 10$^{10}$ &   & -6.29 $\pm$ 1.21& Ski\\
53633 &1.40 $\times$ 10$^{10}$ &   & -4.93 $\pm$ 0.80& Ski\\
53669 &1.50 $\times$ 10$^{10}$ &   & -5.55 $\pm$ 0.51& Ski\\
54350 &1.40 $\times$ 10$^{10}$ &   & -4.91 $\pm$ 0.41& Ski\\
54377 &1.50 $\times$ 10$^{10}$ &   & -5.81 $\pm$ 0.54& Ski\\
54686 &1.60 $\times$ 10$^{10}$ &   & -6.11 $\pm$ 0.21& Ski\\
54690 &1.60 $\times$ 10$^{10}$ &   & -6.38 $\pm$ 0.70& Ski\\
55102 &1.60 $\times$ 10$^{10}$ &   & -6.38 $\pm$ 0.23& Ski\\
55192.01 & 1.80 $\times$ 10$^{10}$ &   & -7.66 $\pm$ 1.27& LT\\
55192.01 & 1.70 $\times$ 10$^{10}$ &   & -6.98 $\pm$ 0.99& LT\\
55207.81 & 2.00 $\times$ 10$^{10}$ &   & -8.90 $\pm$ 1.57& LT\\
55207.82 & 2.10 $\times$ 10$^{10}$ &   & -9.27 $\pm$ 1.27& LT\\
55220.81 & 1.60 $\times$ 10$^{10}$ &   & -6.24 $\pm$ 0.96& LT\\
55220.82 & 2.00 $\times$ 10$^{10}$ &   & -9.05 $\pm$ 0.88& LT\\
55220.82 & 2.20 $\times$ 10$^{10}$ &   & -9.79 $\pm$ 0.47& LT\\
55220.92 & 2.10 $\times$ 10$^{10}$ &   & -9.39 $\pm$ 0.58& LT\\
55220.93 & 2.30 $\times$ 10$^{10}$ &  & -10.60 $\pm$ 0.66& LT\\
55220.93 & 2.00 $\times$ 10$^{10}$ &   & -8.71 $\pm$ 0.95& LT\\
55221.81 & 1.60 $\times$ 10$^{10}$ &   & -6.38 $\pm$ 1.23& LT\\
55221.82 & 2.10 $\times$ 10$^{10}$ &   & -9.48 $\pm$ 0.76& LT\\
55221.83 & 2.30 $\times$ 10$^{10}$ &  & -10.55 $\pm$ 0.65& LT\\
55221.92 & 2.10 $\times$ 10$^{10}$ &   & -9.76 $\pm$ 0.99& LT\\
55221.93 & 2.20 $\times$ 10$^{10}$ &  & -10.08 $\pm$ 0.46& LT\\
55221.94 & 2.30 $\times$ 10$^{10}$ &  & -10.48 $\pm$ 0.43& LT\\
55222.03 & 2.00 $\times$ 10$^{10}$ &   & -9.08 $\pm$ 0.63& LT\\
55222.04 & 2.20 $\times$ 10$^{10}$ &  & -10.35 $\pm$ 0.21& LT\\
55222.05 & 1.90 $\times$ 10$^{10}$ &   & -8.34 $\pm$ 0.32& LT\\
55374.20 & 1.20 $\times$ 10$^{10}$ &   & -4.19 $\pm$ 0.37& LT\\
55374.21 & 1.20 $\times$ 10$^{10}$ &   & -4.08 $\pm$ 0.53& LT\\
55374.21 & 1.20 $\times$ 10$^{10}$ &   & -4.09 $\pm$ 0.59& LT\\
55402.12 & 8.10 $\times$ 10$^{09}$ &  & -2.28 $\pm$ 0.76& LT\\
55402.13 & 7.10 $\times$ 10$^{09}$ &  & -1.86 $\pm$ 0.70& LT\\
55402.14 & 8.60 $\times$ 10$^{09}$ &  & -2.48 $\pm$ 0.73& LT\\
55411 &1.30 $\times$ 10$^{10}$ &   & -4.82 $\pm$ 0.27& Ski\\
55436 &1.40 $\times$ 10$^{10}$ &   & -5.07 $\pm$ 0.11& Ski\\
55453 &1.60 $\times$ 10$^{10}$ &   & -6.19 $\pm$ 0.23& Ski\\
55469 &1.50 $\times$ 10$^{10}$ &   & -5.66 $\pm$ 0.56& Ski\\
55794 &1.70 $\times$ 10$^{10}$ &   & -6.65 $\pm$ 0.87& Ski\\
55810 &2.20 $\times$ 10$^{10}$ &  & -10.13 $\pm$ 0.27& Ski\\
56164 &1.60 $\times$ 10$^{10}$ &   & -6.12 $\pm$ 0.12& Ski\\
56184 &1.70 $\times$ 10$^{10}$ &   & -6.68 $\pm$ 0.65& Ski\\
56535 &1.50 $\times$ 10$^{10}$ &   & -5.79 $\pm$ 0.31& Ski\\
56584 &1.80 $\times$ 10$^{10}$ &   & -7.23 $\pm$ 0.40& Ski\\
  \end{longtable}
  \end{center}
\newpage

   \begin{center}
   
   \begin{longtable}{| c  c  c  c  c |}
   \caption{\textbf{EXO 2030+375: EW(H$_\alpha$) and $\Delta$V measurements}} \\
   \hline
   \multicolumn{1}{|c }{{MJD}} & \multicolumn{1}{c}{{radius (m)}} & \multicolumn{1}{c}{{$\Delta$V (km/s)}} & \multicolumn{1}{c}{{EW(H$_{\alpha}$) (\AA) }} & \multicolumn{1}{c|}{{Telescope }}\\ \hline 
\endfirsthead

\multicolumn{3}{c}%
{{\bfseries \tablename\ \thetable{} -- continued from previous page}} \\
\hline
 \multicolumn{1}{|c }{{MJD}} & \multicolumn{1}{c}{{radius (m)}} & \multicolumn{1}{c}{{$\Delta$V (km/s)}} & \multicolumn{1}{c}{{EW(H$_{\alpha}$) (\AA) }} & \multicolumn{1}{c|}{{Telescope }}\\ \hline 
\endhead

\hline \multicolumn{3}{r}{{Continued on next page}} \\ \hline
\endfoot

\hline \hline
\endlastfoot

    \hline
46680 & 1.71 $\times$ 10$^{11}$ &  & -15 $\pm$ 0.95& INT\\
48897 & 2.56 $\times$ 10$^{11}$ &  & -27.32 $\pm$ 1.78& WHT\\
48898 & 2.32 $\times$ 10$^{11}$ &  & -23.62 $\pm$ 1.54& WHT\\
48898 & 2.33 $\times$ 10$^{11}$ &  & -23.75 $\pm$ 1.54& WHT\\
49166 & 1.84 $\times$ 10$^{11}$ &  & -16.77 $\pm$ 1.00& WHT\\
49166 & 1.97 $\times$ 10$^{11}$ &  & -18.54 $\pm$ 1.21& WHT\\
49166 & 1.92 $\times$ 10$^{11}$ &  & -17.78 $\pm$ 1.16& WHT\\
49167 & 1.71 $\times$ 10$^{11}$ &  & -15.04 $\pm$ 0.98& WHT\\
49167 & 1.62 $\times$ 10$^{11}$ &  & -13.80 $\pm$ 0.90& WHT\\
49167 & 1.91 $\times$ 10$^{11}$ &  & -17.68 $\pm$ 1.15& WHT\\
49167 & 2.15 $\times$ 10$^{11}$ &  & -21.05 $\pm$ 1.37& WHT\\
49167 & 1.95 $\times$ 10$^{11}$ &  & -18.27 $\pm$ 1.19& WHT\\
49167 & 1.85 $\times$ 10$^{11}$ &  & -16.85 $\pm$ 1.10& WHT\\
49167 & 1.84 $\times$ 10$^{11}$ &  & -16.72 $\pm$ 1.09& WHT\\
49167 & 1.96 $\times$ 10$^{11}$ &  & -18.41 $\pm$ 1.20& WHT\\
49167 & 2.09 $\times$ 10$^{11}$ &  & -20.27 $\pm$ 0.70& WHT\\
49167 & 2.15 $\times$ 10$^{11}$ &  & -21.09 $\pm$ 1.37& WHT\\
49167 & 2.06 $\times$ 10$^{11}$ &  & -19.74 $\pm$ 1.10& WHT\\
49167 & 2.22 $\times$ 10$^{11}$ &  & -22.14 $\pm$ 0.9& WHT\\
49168 & 1.65 $\times$ 10$^{11}$ &  & -14.21 $\pm$ 0.67& WHT\\
49168 & 1.67 $\times$ 10$^{11}$ &  & -14.45 $\pm$ 1.17& WHT\\
49168 & 2.02 $\times$ 10$^{11}$ &  & -19.20 $\pm$ 1.29& WHT\\
49168 & 2.08 $\times$ 10$^{11}$ &  & -20.09 $\pm$ 1.21& WHT\\
49168 & 1.76 $\times$ 10$^{11}$ &  & -15.59 $\pm$ 0.64& WHT\\
49168 & 1.86 $\times$ 10$^{11}$ &  & -16.96 $\pm$ 0.52& WHT\\
49168 & 1.94 $\times$ 10$^{11}$ &  & -18.09 $\pm$ 0.78& WHT\\
49168 & 2.07 $\times$ 10$^{11}$ &  & -19.90 $\pm$ 0.77& WHT\\
49168 & 1.98 $\times$ 10$^{11}$ &  & -18.64 $\pm$ 1.21& WHT\\
50272 & 1.30 $\times$ 10$^{11}$ &   & -9.88 $\pm$ 1.30& WHT\\
50273 & 1.20 $\times$ 10$^{11}$ &   & -8.80 $\pm$ 0.87& WHT\\
50273 & 1.50 $\times$ 10$^{11}$ &  & -12.29 $\pm$ 1.43& WHT\\
50273 & 1.48 $\times$ 10$^{11}$ &  & -12.11 $\pm$ 1.90& WHT\\
54004 & 1.19 $\times$ 10$^{11}$ &   & -8.70 $\pm$ 0.70 & RT\\
54093 & 1.21 $\times$ 10$^{11}$ &   & -8.90 $\pm$ 0.71 & RT\\
54356 & 1.32 $\times$ 10$^{11}$ &  & -10.20 $\pm$ 0.82 & RT\\
  \end{longtable}
  \end{center}

\newpage

   \begin{center}
   
   \begin{longtable}{| c  c  c  c  c |}
   \caption{\textbf{1A~1118$-$61: EW(H$_\alpha$) and $\Delta$V measurements }} \\
   \hline
   \multicolumn{1}{|c }{{MJD}} & \multicolumn{1}{c}{{radius (m)}} & \multicolumn{1}{c}{{$\Delta$V (m/s)}} & \multicolumn{1}{c}{{EW(H$_{\alpha}$) (\AA) }} & \multicolumn{1}{c|}{{Telescope }}\\ \hline 
\endfirsthead

\multicolumn{3}{c}%
{{\bfseries \tablename\ \thetable{} -- continued from previous page}} \\
\hline
 \multicolumn{1}{|c }{{MJD}} & \multicolumn{1}{c}{{radius (m)}} & \multicolumn{1}{c}{{$\Delta$V (m/s)}} & \multicolumn{1}{c}{{EW(H$_{\alpha}$) (\AA) }} & \multicolumn{1}{c|}{{Telescope }}\\ \hline 
\endhead

\hline \multicolumn{3}{r}{{Continued on next page}} \\ \hline
\endfoot

\hline \hline
\endlastfoot

    \hline

49049 & 8.92 $\times$ 10$^{10}$ &  & -64.54 $\pm$ 3.58& SAAO\\
49050 & 9.36 $\times$ 10$^{10}$ &  & -69.27 $\pm$ 2.91& SAAO\\
49160 & 9.25 $\times$ 10$^{10}$ &  & -68.10 $\pm$ 0.97& SAAO\\
49161 & 8.35 $\times$ 10$^{10}$ &  & -58.41 $\pm$ 3.83& SAAO\\
49418 & 8.71 $\times$ 10$^{10}$ &  & -62.28 $\pm$ 3.21& SAAO\\
49535 & 8.62 $\times$ 10$^{10}$ &  & -61.32 $\pm$ 1.98& SAAO\\
49536 & 8.64 $\times$ 10$^{10}$ &  & -61.53 $\pm$ 3.91& SAAO\\
49773 & 9.16 $\times$ 10$^{10}$ &  & -67.06 $\pm$ 2.17& SAAO\\
50176 & 8.77 $\times$ 10$^{10}$ &  & -62.84 $\pm$ 2.73& SAAO\\
50619 & 9.54 $\times$ 10$^{10}$ &  & -71.29 $\pm$ 0.95& SAAO\\
50849 & 8.68 $\times$ 10$^{10}$ &  & -61.88 $\pm$ 2.47& SAAO\\
51187 & 8.63 $\times$ 10$^{10}$ &  & -61.39 $\pm$ 2.54& SAAO\\
53444 & 9.05 $\times$ 10$^{10}$ &  & -65.95 $\pm$ 2.51& SAAO\\
53448 & 9.31 $\times$ 10$^{10}$ &  & -68.73 $\pm$ 3.31& SAAO\\
54226 & 9.01 $\times$ 10$^{10}$ &  & -65.42 $\pm$ 3.67& SAAO\\
56317.46 & 8.75 $\times$ 10$^{10}$ &  & -62.69 $\pm$ 2.92&SALT\\
56350.60 & 8.66 $\times$ 10$^{10}$ &  & -61.73 $\pm$ 2.66&SALT\\
  \end{longtable}
  \end{center}

\end{document}